\documentclass[a4paper,fleqn,usenatbib]{mnras}
\pdfoutput=1

\usepackage[T1]{fontenc}
\usepackage{ae,aecompl}

\usepackage{pdflscape}
\usepackage{appendix}
\usepackage{graphicx}	
\usepackage{amsmath}	
\usepackage{amssymb}	
\usepackage[labelfont=bf]{subcaption}

\makeatletter
\newcommand*{\rom}[1]{\expandafter\@slowromancap\romannumeral #1@}
\makeatother

\title[An Improved Age-Activity Relationship]{An Improved Age-Activity Relationship for Cool Stars older than a Gigayear}

\author[R. S. Booth, K. Poppenhaeger et al.]{
{R. S. Booth$^{1}$\thanks{rbooth03@qub.ac.uk}, K. Poppenhaeger$^{1,2}$, C. A. Watson$^{1}$, V. Silva Aguirre$^{3}$, and S. J. Wolk$^{2}$}
\\
$^{1}$Astrophysics Research Centre, School of Mathematics \& Physics, Queen's University Belfast, University Road, Belfast BT7 1NN, UK\\
$^{2}$Harvard-Smithsonian Center for Astrophysics, 60 Garden Street, Cambridge, 02138 MA, USA\\
$^{3}$Stellar Astrophysics Centre, Department of Physics and Astronomy, Aarhus University, Ny Munkegade 120, DK-8000 Aaarhus C, Denmark
}

\date{Accepted XXX. Received YYY; in original form ZZZ}

\pubyear{2017}

\begin{document}
\label{firstpage}
\pagerange{\pageref{firstpage}--\pageref{lastpage}}
\maketitle

\begin{abstract}
Stars with convective envelopes display magnetic activity, which decreases over time due to the magnetic braking of the star. This age-dependence of magnetic activity is well-studied for younger stars, but the nature of this dependence for older stars is not well understood. This is mainly because absolute stellar ages for older stars are hard to measure. However, relatively accurate stellar ages have recently come into reach through asteroseismology. In this work we present X-ray luminosities, which are a measure for magnetic activity displayed by the stellar coronae, for 24 stars with well-determined ages older than a gigayear. We find 14 stars with detectable X-ray luminosities and use these to calibrate the age-activity relationship. We find a relationship between stellar X-ray luminosity, normalized by stellar surface area, and age that is steeper than the relationships found for younger stars, with an exponent of $-2.80 \pm 0.72$. Previous studies have found values for the exponent of the age-activity relationship ranging between $-1.09$ to $-1.40$, dependent on spectral type, for younger stars. Given that there are recent reports of a flattening relationship between age and rotational period for old cool stars, one possible explanation is that we witness a strong steepening of the relationship between activity and rotation.
\end{abstract}


\section{Introduction}
\label{section1}

`Magnetic activity' is a collective term to describe a plethora of magnetically-driven phenomena observed mainly for cool stars, such as flares, coronal mass ejections, starspots, and the very existence of a chromosphere and corona, which are much hotter than the underlying stellar photosphere. These phenomena are caused by the highly localised and time-variable magnetic fields, which are in turn driven by the (radial and longitudinal) differential rotation of these stars. Over time, these stars spin down and magnetic activity becomes less pronounced; observationally, this has led to many studies concerning the evolution of stellar rotation with age \citep{Bouvier1997, Herbst2007} and stellar activity with age \citep{Guedel1997, Preibisch2005, 2005ApJ...622..653T}. \citet{1962AnAp...25...18S} first proposed that angular momentum was removed from stars via magnetic braking where the material lost from the stellar surface (i.e the stellar wind) is kept in corotation with the stellar surface by the magnetic field up to a critical distance. Beyond this the material is lost from the star and carries away some angular momentum. Over time, the angular momentum lost decreases the rotational velocity of the star resulting in a longer stellar rotation period. Magnetic braking needs a substantial magnetic field, which is found in low mass main sequence stars that have a radiative core surrounded by a convective envelope. This causes a dynamo effect which is driven by the interplay between convection and differential rotation \citep{parker_55}. This generates the necessary magnetic fields.

There have been many studies on how stellar rotation varies with age, also known as gyrochronology \citep{barnes_03}. Some studies have used stars with known ages to calibrate the relationship (e.g. \citealt{2007ApJ...669.1167B,mamjek_08,Barnes:2010bf,2015MNRAS.450.1787A}), while others report on stellar data to validate certain gyrochronology relations (e.g. \citealt{Meibom:2011eu,Meibom:2015if,2016ApJ...823...16B}). Semi-physical spin-down models have also been presented in the literature \citep{2013ApJ...776...67V,2015ApJ...799L..23M}. Studies that calibrate the relationship generally construct a power law that incorporates the age, rotation and some mass dependent parameter such as colour or convective turnover time. However, the majority of these studies concentrate on measuring the period of rotation of stars in clusters with well known ages and using these to calibrate the relationship. Recently, \citet{2016Natur.529..181V} measured the rotation periods of older stars with ages determined by asteroseismology, and presented models that incorporated dramatically weakened magnetic braking for these older stars. This reported result may have significant implications for gyrochronology as it calls into question the use of stellar spin-down as an age diagnostic past a given age and would limit the sample of stars that the method could be applied to.

The relationship between magnetic activity and stellar age was first presented by \citet{skumanich_72}. Skumanich plotted the Calcium \rom{2} emission (a proxy for magnetic activity), average equatorial velocities, and lithium abundance as a function of age and noted that the calcium emission and average equatorial velocities followed a similar relationship. Skumanich suggested that this was a consequence of the stellar dynamo; over time the slower stellar rotation leads to a reduced velocity shear at the tachocline between the radiative and convective zones, resulting in reduced magnetic field generation by the stellar dynamo \citep{9780511546037}.

\begin{table*}
	\caption{Table of Observations that were fully analysed and presented in the Results Section. $^*$A large number of observations exist for these targets and has been taken into account in our analysis; one specific observation is listed to illustrative typical exposure times.}
	\centering
	\begin{tabular}{lllc}
		\hline \hline
		Name of Star / Name of White Dwarf & Telescope / Instrument     & Observation ID & Exposure Time / $10^{3}$ s \\
		\hline
		16 Cyg A & Chandra ACIS-I  & 16647 & 35.33 \\
		 & Chandra ACIS-I   & 18756 & 38.57 \\
		16 Cyg B & Chandra ACIS-I  & 16647 & 35.33 \\
		 & Chandra ACIS-I  & 18756 & 38.57 \\
		40 Eri A/40 Eri B & Chandra ACIS-S  & 13644  & 5.00 \\
		61 Cyg A & XMM PN Thick Filter & 41740301$^*$ & 7.09  \\
		61 Cyg B  & XMM PN Thick Filter & 41740301$^*$ & 7.09 \\
		CD -3710500 / L481-60 & Chandra ACIS-I   & 13769 & 24.75 \\
		GJ 176 & Chandra ACIS-S  & 13638 & 4.98  \\
		GJ 191 & Chandra ACIS-S   & 13646 & 5.00 \\
		HR 7703 & XMM PN Thick Filter & 670380401 & 7.95  \\
		KIC 10016239 & XMM PN Medium Filter & 761560601 & 22.24  \\
		KIC 12011630 & Chandra ACIS-I HETG  & 9969 & 19.74 \\
		KIC 3123191 & Chandra ACIS-I  & 13166 & 2.72 \\
		KIC 5309966 & Chandra ACIS-I   & 17138 & 26.43 \\
		 & Chandra ACIS-I  & 17141 & 29.69  \\
		 & Chandra ACIS-I  & 17513 & 49.40 \\
		 & Chandra ACIS-I  & 17516  & 49.02  \\
		KIC 6116048 & XMM PN Medium Filter & 722330201 & 11.58 \\
		 & XMM PN Medium Filter & 722330301 & 10.73 \\
		KIC 6603624 & XMM PN Medium Filter & 761560701 & 35.67 \\
		KIC 7529180  & XMM PN Thin Filter & 743460201 & 18.07 \\
		KIC 8292840 & XMM PN Medium Filter & 743840201 & 10.74 \\
		KIC 9025370 & XMM PN Medium Filter & 761560501 & 8.92 \\
		KIC 9410862 & XMM PN Medium Filter & 670140501 & 26.36\\
		KIC 9955598 & XMM PN Medium Filter & 761560601 & 22.24 \\
		NLTT 7887 / NLTT 7890 & XMM PN Medium Filter & 670650101 & 16.60  \\
		Proxima Centauri & XMM PN Medium Filter & 551120201$^*$ & 26.49 \\
		Alpha Centauri B & XMM PN Thick Filter & 760290301$^*$ & 14.88 \\     
		\hline
		\end{tabular}
		\label{table:obs}
\end{table*}

Several proxies for magnetic activity have been studied in the literature. Among them is the emission in chromospheric lines such as Ca II H \& K \citep{pace_cahk_age,1984ApJ...279..763N} as well as H-$\alpha$. A study of the chromospheric emission usually needs to include an analysis of the so-called basal part of the emission in the relevant lines, i.e. the part of the line emission that is not driven by magnetic activity \citep{Schrijver1987}. However, the analysis of the basal flux does not account for metallicity or surface gravity, which has an effect on the shape of the Ca II profiles \citep{1998MNRAS.298..332R}. This makes calibrating the age-activity relationship using Ca II emission more complex (e.g. \citealt{2016A&A...594L...3L}). Calcium emission is also particularly difficult to study for M dwarfs due to the low level of continuum present in the region around the Ca II H \& K lines. Consequently, calcium emission studies tend to focus on FGK stars and the H-$\alpha$ line is widely used as an activity indicator in M dwarfs (e.g. \citealt{2017ApJ...834...85N,2012AJ....143...93R}).

Another proxy for magnetic activity is the emission from the stellar corona \citep{Erdelyi:2007gw, 2001A&A...377..538S,Preibisch2005,jackson_lx_age}, which is mainly observed in the soft X-ray band (0.1-10~keV). This emission stems from the hot (several million Kelvin) plasma in the stellar corona, which is collisionally excited and cools through emission at X-ray wavelengths \citep{Raymond1977}. Since the X-ray luminosity does not contain a photospheric component, it means that the X-ray luminosity is unambiguously associated with magnetic heating unlike Ca II H \& K emission. In this work we use the X-ray luminosity as a magnetic activity indicator in order to study the full range of stars with convective envelopes (mid-F to mid-M).

To study the evolution of stellar activity for old stars, it is crucial to obtain a good estimate for the stellar age in the first place. This can be difficult for stars older than a gigayear, as most age-determination methods work best for younger stars. However, recent observational advances have made it possible to study ages for a larger number of stars through asteroseismology. Asteroseismology provides critical information about the interior of stars through observations of stellar oscillations. This has become a valuable tool since the launch of the CoRoT and Kepler missions, which have provided higher quality photometry \citep{chaplin_astero} enabling accurate and precise measurements of  fundamental properties of stars, including ages \citep{Metcalfe:2012kv,Mathur:2012bj,astero2,astero1}. Indeed, asteroseismology has proved to be the most accurate age-dating method for old field stars -- and opens up the possibility of stellar age investigations for stars older than a gigayear.

The age-activity relationship is not only useful for inferring ages for stars with $L_{x}$ measurements, but it can also give us insight into the high-energy environment that stars provide their exoplanets \citep{2004Icar..170..167Y,2012A&A...541A..26P}, and how this changes over time. This is important when considering the effects of high-energy radiation on the habitability of exoplanets.

In this work, we use ages from recent asteroseismology studies coupled to X-ray luminosities for these stars to investigate the age-activity relationship for stars older than a gigayear. In Section \ref{section2} we present the data used in the analysis followed by Section \ref{section3} which details the analysis performed on the data. Section \ref{section4} presents the results and Section \ref{section5} presents the discussion. Finally, Section \ref{section6} summarises the conclusions from this work.

\section{Observations}
\label{section2}

\subsection{Sample selection}
\label{sample_section}

Our target stars with well-determined old ages were selected from different sources. The majority stems from asteroseismology where we chose stars with precisely determined stellar properties, including ages, obtained by modelling the individual oscillation frequencies in the spectrum observed by the \textit{Kepler} satellite \citep{astero1,SilvaAguirre2017}. For stars where the \textit{Kepler} observations are not of sufficient signal-to-noise ratio to extract the individual oscillation modes, we combined the asteroseismic detections reported by \citealt{astero2} with the spectroscopic effective temperatures and metallicities derived by \citealt{2015ApJ...808..187B} and determined asteroseismic ages using the BAyesian STellar Algorithm (BASTA, \citealt{astero1}). Specifically, this new age determination was performed for the stars KIC 10016239, KIC 12011630, KIC 3123191, KIC 5309966, and KIC 7529180, yielding asteroseismically determined stellar ages and radii. Near-by stars were selected for dedicated X-ray observations with \textit{XMM-Newton} and \textit{Chandra} (PI Poppenhaeger).
The second source was a sample of G and K type stars with archival X-ray observations that are located in wide binaries containing a white dwarf companion. Targets with existing X-ray observations were identified from the samples in \citet{garces_wd} and \citet{2012ApJ...746..144Z}. These are particularly useful systems as the ages of the white dwarfs are reasonably well known through their cooling times and therefore can act as a stellar chronometer for the system that is independent of spin-down. We have calculated the ages of those systems from the white dwarf parameters, which we explain in more detail in Section \ref{wd_ages}.
The third source were individually selected stars with archival X-ray observations and relatively well known ages determined through various methods. These methods included asteroseismology (16 Cyg A and B, \citealp{SilvaAguirre2017}; the $\alpha$ Cen/Proxima Cen system, \citealp{prox_cen_age_2}), isochrone fitting (61 Cyg A and B, \citealp{61_cyg_age}) and association with a sub-population of stars in the galaxy (HR7703, \citealp{2010_dewarf}). Proxima Centauri is a fully convective star, so one might wonder if it is appropriate to include in this sample of otherwise cool stars with radiative cores; however, a recent study \citep{2016Natur.535..526W} found that fully convective stars exhibit a rotation-activity relationship that is indistinguishable to that of solar-type stars, which is why we chose to include Proxima Cen in our analysis. Also included in our sample is the Sun; its X-ray luminosity was adapted using the model parameters in \citet{2000ApJ...528..537P} and Xspec to encompass only the 0.2 - 2.0 keV energy range. The solar age is well constrained by meteorite studies and is adopted to be $4.57 \pm 0.02$ Gyr \citep{1995RvMP...67..781B}.

For stars with exoplanets in close orbits, effects on the stellar activity through star-planet interaction are expected from theoretical considerations \citep{Cuntz2000} and have been observed for some systems with high-mass exoplanets in very close orbits (see for example \citealt{Poppenhaeger2014, Pillitteri2015}). However, in our sample there are no stars with Hot Jupiters present, therefore such effects are not expected to play a role in our investigation. 

Two X-ray observatories are used in our study, \textit{XMM-Newton} and \textit{Chandra}. Their main characteristics and the basics of our data reduction are shortly explained in the following paragraphs. The details of the observations (obtained from both \textit{XMM-Newton} and \textit{Chandra}) that we analysed are presented in Section \ref{section4} and listed in Table \ref{table:obs}.

The \textit{XMM-Newton} X-ray Telescope \citep{Jansen2001} is equipped with the European Photon Imaging Camera (EPIC) which consists of three X-ray CCD cameras, one PN and two metal oxide semi-conductor (MOS) cameras. EPIC allows imaging over the telescope's 30 arc-minute field of view and in the energy range of 0.2 to 15 keV making it suitable to study the X-ray emission from late-type stars. Archival observations were obtained through the XMM-Newton Science Archive and analysed using SAS (Science Analysis System) version 15.0. Using SAS, we filtered the data to remove any bad pixels or bad events by setting criteria to limit the probability of double photon impact events. These events occur when two photons hit the same or neighbouring pixels in the same readout time frame and cause a slightly different pattern on the chip compared to a single photon event. In our data analysis of the observations from \textit{XMM-Newton} we used a standard source extraction radius of 20" and chose a source-free background region with a radius of 70". Data analysis was preferably performed with PN observations due to the higher signal to noise obtained with this instrument.

The \textit{Chandra} X-ray telescope \citep{Weisskopf2000} has two focal plane instruments, the Advanced CCD Imaging Spectrometer (ACIS) and the High Resolution camera (HRC). The ACIS provides images alongside spectral information on the object in the energy range 0.2 - 10 keV. The HRC only provides images and no spectrally resolved data; we therefore restricted our analysis to observations conducted with ACIS. Observation files from \textit{Chandra} were obtained through the Chandra X-ray Centre (CXC) public archive and were analysed with CIAO (Chandra Interactive Analysis of Observations) \citep{ciao}. We used a standard source extraction radius of 1.5" and a source-free background region with a radius of 15".

\subsection{Distances and spectral types}
\label{dist_section}

Since the rotational evolution and the X-ray luminosity of stars no longer on the main sequence differs from those still on it, it was important to ensure only main sequence stars were considered. For several stars from the asteroseismic part of the sample no luminosity classes were given in the literature. We therefore compared their surface gravity to the relation of $B-V$ colour and surface gravity for main-sequence stars, as given by \cite{CBO9781316036570A207}, and excluded stars which differed by more than 0.2~dex from our sample. 

For one of the stars from the asteroseismic part of our sample, no distances or parallaxes were found in the literature. Therefore, for this star the Barnes-Evans method was used. This method was used to calculate the angular diameter of the star using $V-K$ \citep{1997A&A...320..799F}. Since the radius of the star is known from asteroseismology \citep{astero2,2015ApJ...808..187B} it was then possible to calculate the distance to the star. As the stars in this study are all located relatively nearby, reddening is expected to be insignificant and was not taken into account. The Barnes-Evans method has been used previously to obtain stellar radii for extra-solar planet host stars and found good agreement with published values \citep{2010MNRAS.408.1606W}. The X-ray luminosity was then calculated using the flux (see Section \ref{xray_flux}) and distance determined from the appropriate method.

Finally it was necessary to determine the spectral type of the candidates since this influences both the stellar activity and its evolution with time. The stellar spectral types were collected from the SIMBAD database, or estimated from the stellar effective temperatures as published in \citealt{astero2,astero1}.

\subsection{Ages of systems with white dwarfs}
\label{wd_ages}

Cool stars located in a wide binary system with a white dwarf provide the opportunity to infer the age of the system and therefore the age of the cool star from the physical properties of the white dwarf. When the effective temperature and surface gravity of the white dwarf are known, one can infer its mass and cooling time, and estimate the mass of the progenitor star and its main sequence lifetime. The sum of the main sequence lifetime and the white dwarf cooling time is then the total age of the system.

\citet{garces_wd} have performed such age estimates for several systems; however, investigations into 3D-model atmospheres of white dwarfs revealed that previous fits of spectra to 1D-models overestimated the true effective temperatures $T_{\mathrm{eff}}$ and surface gravities $\log g$ of cool ($<13000$~K) white dwarfs \citep{Tremblay2013}. We have therefore re-calculated the ages for the systems we use in this work, specifically the systems 40 Eri A/40 Eri B, CD -3710500/L481-60, and NLTT 7887/NLTT 7890, where the second object in each listed pair is the white dwarf. Specifically, we used the published $T_{\mathrm{eff}}$ and $\log g$ values from \citet{garces_wd} and \citet{2012ApJ...746..144Z} for the white dwarfs and corrected them according to the formulae (7) and (8) from \citet{Tremblay2013}. We then used the Montreal model grids for white dwarfs with hydrogen atmospheres\footnote{Available at \url{http://www.astro.umontreal.ca/~bergeron/CoolingModels/}} to estimate the masses and cooling ages of the white dwarfs \citep{Holberg2006, Kowalski2006, Tremblay2011, Bergeron2011}; progenitor masses were estimated using the initial-final mass relation by \citet{2008ApJ...676..594K}, and the progenitor main-sequence lifetimes were estimated using the Padova stellar evolution model grids from \citet{Bertelli2008}. In this manner, we estimate the system ages for the 40 Eri A/40 Eri B and CD -3710500/L481-60 systems to be $3.70^{+3.57}_{-1.34}$~Gyr and $1.77^{+0.65}_{-0.27}$~Gyr, respectively. NLTT 7890 has large error bars on its surface gravity given by \citet{garces_wd}, therefore the estimated system age for NLTT 7887/NLTT 7890 has larger errors with $4.97^{+8.8}_{-3.0}$~Gyr.

\begin{table*}
	\caption{Conversion factors used by WebPIMMS to convert number of counts into flux in the 0.2 - 2.0 keV energy range, assuming a coronal temperature of $\log T = 6.5$ and solar abundances.}
	\centering
	\begin{tabular}{l l }
		\hline \hline
		Name of Instrument & Conversion Factor \\
         & (erg s$^{-1}$ cm$^{-2}$ count$^{-1}$) \\ 
		\hline
		XMM PN Medium Filter & $1.03\times10^{-12}$\\
		Chandra ACIS-I No Grating (Cycle 16) & $2.42\times10^{-11}$\\
		Chandra ACIS-I No Grating (Cycle 12) & $1.52\times10^{-11}$\\
		Chandra ACIS-I HET Grating (Cycle 10) & $2.16\times10^{-10}$\\
		\hline
	\end{tabular}
	\label{table:webpimms}
\end{table*}

\section{Data Analysis}
\label{section3}

In the following section we will provide details on the methods used to determine the X-ray luminosity for our sample of stars. In addition, Appendix \ref{additional_info} provides supplementary information on the analysis of individual stars that are not included in this section.

\subsection{Source detections}
For each of our targets we tested if the source was significantly detected in X-rays. We extracted X-ray counts in the energy band from 0.2-2 keV, as this is where weakly active cool stars display most of their X-ray emission, for the background and source regions. 

For \textit{XMM-Newton}, there is typically a significant background signal observed in any observation. We therefore tested for \textit{XMM-Newton} targets if the number of source counts exceeded the number of counts expected form a pure background signal (estimated from the larger background region) by at least $3\sigma$, with $\sigma$ being estimated as the square root of the number of expected background counts. If so, we counted the source as detected and proceeded with a flux determination (see Section \ref{xray_flux}); otherwise, we chose the $3\sigma$ level over the background signal as the upper limit for the source.

For \textit{Chandra}, the background signal is typically very low, meaning that an approximation of using the square root as the error on the expected background counts is invalid. We therefore used full Poisson statistics and calculated the inverse percent point function at which, for a given expected number of background counts, the number of counts in the source region had a probability of less than 0.3\% of occurring as a random fluctuation. If the number of observed source counts was at this number or larger, we counted the source as detected; otherwise we used that number as the upper limit on the X-ray counts for the source. 

\subsection{Stellar Variability}
\label{variability}
Magnetic activity can vary on several timescales, including timescales shorter than the typical exposure time for an X-ray observation. Therefore, for detected target stars, a light curve was extracted from each observation to check for short term magnetic phenomena such as flares. In the case of Proxima Centauri, the light curve showed several rapid increases in count rate over the observation timescale, indicating several flares. Flares increase the temperature and emission  measure of the corona, resulting in a significantly higher X-ray luminosity compared to the quiescent emission level of the star. Therefore, the quiescent value for the X-ray luminosity of $4.9 \times 10^{26}$ ergs s$^{-1}$ was taken from \citet{2011A&A...534A.133F} and used in this work. An inspection of the light curve for HR 7703 indicated that a flare had occurred towards the end of the observation. In this case, the time interval associated with this flare was excluded from the data analysis.

\begin{figure*}
\includegraphics[scale=0.41]{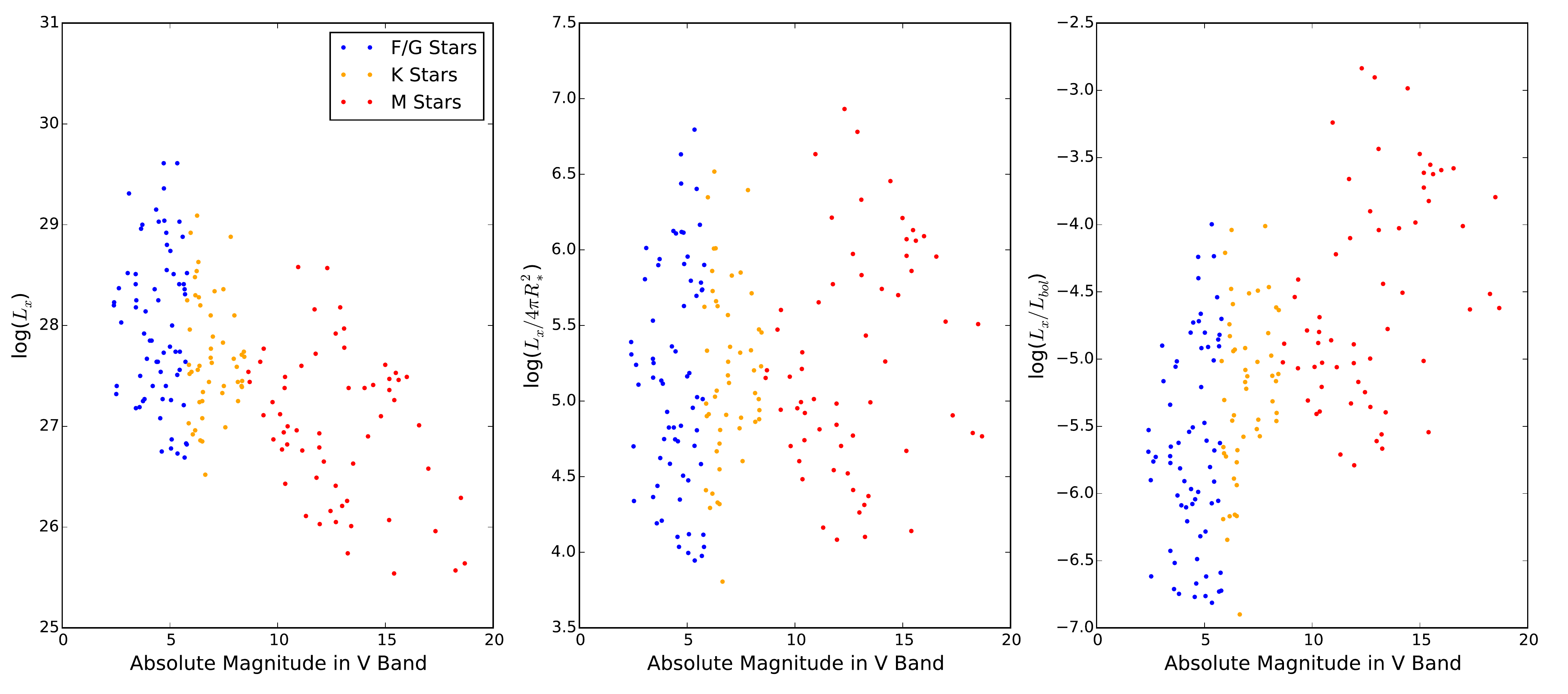}
\caption{Comparison of data taken from \citet{2004A&A...417..651S} when normalized by bolometric luminosity and stellar surface. \textbf{Left Panel}: Logarithmic value of X-ray luminosity as a function of absolute magnitude in V band. \textbf{Middle Panel}: Logaritmic value of X-ray luminosity normalized by the stellar surface area as a function of absolute magnitude in V band. \textbf{Right Panel}: Logaritmic value of the ratio of X-ray luminosity to bolometric luminosity as a function of absolute magnitude in V band.}
\centering
\label{schmitt_data}
\end{figure*}

\subsection{Determining the X-ray Flux}
\label{xray_flux}
For X-ray sources with significant detections and where the source region contained $\approx 90$ or more counts, we chose to model the spectra of the source with a coronal plasma model using the following method. A spectrum of the star was extracted from the observation using the relevant analysis tools for each telescope. The extracted spectra were fitted with an optically thin thermal plasma model (APEC model) using the Xspec fitting software. Since all of the stars in this study are located nearby the redshift was fixed at zero; the abundances were assumed to be solar using abundances from \citet{1998SSRv...85..161G}. The two variables that were fitted were coronal temperature and emission measure; in some cases, two temperature components were required to find a fit that describes the coronal emission well. From the best fit model to each object the flux was then calculated in a fixed energy band from 0.2-2~keV. The details and plots of the best fit models are shown in Appendix \ref{spec_model_results} and Appendix \ref{spec_model_plots} respectively.

In the case where the source region contained less than $\approx 90$ counts then typically there were not enough data points to fit a spectrum accurately. Under these circumstances an estimate of the X-ray flux was obtained through WebPIMMS\footnote{\url{https://heasarc.gsfc.nasa.gov/cgi-bin/Tools/w3pimms/w3pimms.pl}} using the mean count rate of the source region. A typical spectrum was assumed for the stellar corona and WebPIMMS calculated the source flux using the instrument characteristics. The X-ray flux was calculated in the 0.2 keV and 2.0 keV energy range assuming an Apec model of solar abundance and $log$T value of 6.5 ($T\approx 3$~MK), appropriate for inactive cool stars (see for example \citealt{2005ApJ...622..653T, Johnstone2015}). The conversion factor from counts to flux used for each of the instruments are shown in Table \ref{table:webpimms}; note that the sensitivity of \textit{Chandra} changes significantly over the years of operation, and therefore the correct conversion factors need to be chosen for the observing cycle in which a given observation took place. For \textit{XMM-Newton} observations, the encircled energy fraction factor needs to be applied as well, since its PSF is significantly larger than typical practical source extraction radii.

The statistical error on the values of X-ray counts were calculated from the square root of the number of source counts ($C_{s}$) (as the distribution is described by Poisson statistics) and divided by the number of source counts to obtain a fractional error. This fractional error on the number of counts was also used as the error on the X-ray flux and the X-ray luminosity. However, the X-ray luminosity of a star is known to vary on various timescales, including timescales much shorter than the star's main sequence lifetime, therefore a minimum physical error of 0.1 dex in $\log L_{x}$ was applied to the data to account for this variability, even if the statistical error was smaller. This value was determined from the long-term X-ray monitoring of 61 Cyg B \citep{2012A&A...543A..84R}, a star without an apparent activity cycle, where the standard deviation of the X-ray luminosity was at 0.1 dex over several years of observations.

\section{Results}
\label{section4}

\subsection{Magnetic activity across spectral types}
It is known that there is a mass dependence on the rotational spin down of late type stars due to the varying depth of the convection zone. F type stars have a thinner convection zone resulting in rotational spin down that occurs on a different timescale than for M type stars that have much thicker convection zones. This mass dependency seen in the rotational spin down is also present in the X-ray activity across varying spectral types. Since the sample of our stars is relatively small, we wish to avoid splitting the sample by spectral type and perform an activity-age analysis of the whole sample instead.

When dealing with X-ray activity, some studies normalize $L_X$ by the stellar bolometric luminosity and then split the sample into different mass bins \citep{jackson_lx_age,Preibisch2005}. A different approach was demonstrated by \citet{2004A&A...417..651S}: in their volume-complete sample of cool stars in the solar neighbourhood, they found that when the X-ray luminosity is divided by the stellar surface area $4\pi R_\ast^2$ with $R_\ast$ being the stellar radius, i.e.\ when one considers the \textit{X-ray flux through the stellar surface} as a quantity of interest, then stars of all spectral types from F to M show the same spread of this quantity. We visualise the relevant quantities in Fig.~\ref{schmitt_data}. We took the X-ray luminosities of near-by stars reported by \citep{2004A&A...417..651S}, giving preference to data collected with the \textit{PSPC} detector without the Boron filter if several detections were reported, and plot them against absolute stellar brightness as a proxy for spectral type. We also show the X-ray flux through the stellar surface, as well as the X-ray luminosity divided by the bolometric luminosity. As can be clearly seen, a flat distribution of the X-ray surface flux versus spectral type is present, as has been reported by \citet{2004A&A...417..651S}.

We therefore chose to follow this approach and use a normalization by stellar surface in our data in order to perform a combined analysis of all spectral types present in our sample. Stellar radii were either taken from the asteroseismic studies as mentioned in Section \ref{sample_section}, or calculated from absolute brightnesses and stellar effective temperatures.

\begin{figure*}
\includegraphics[scale=0.65]{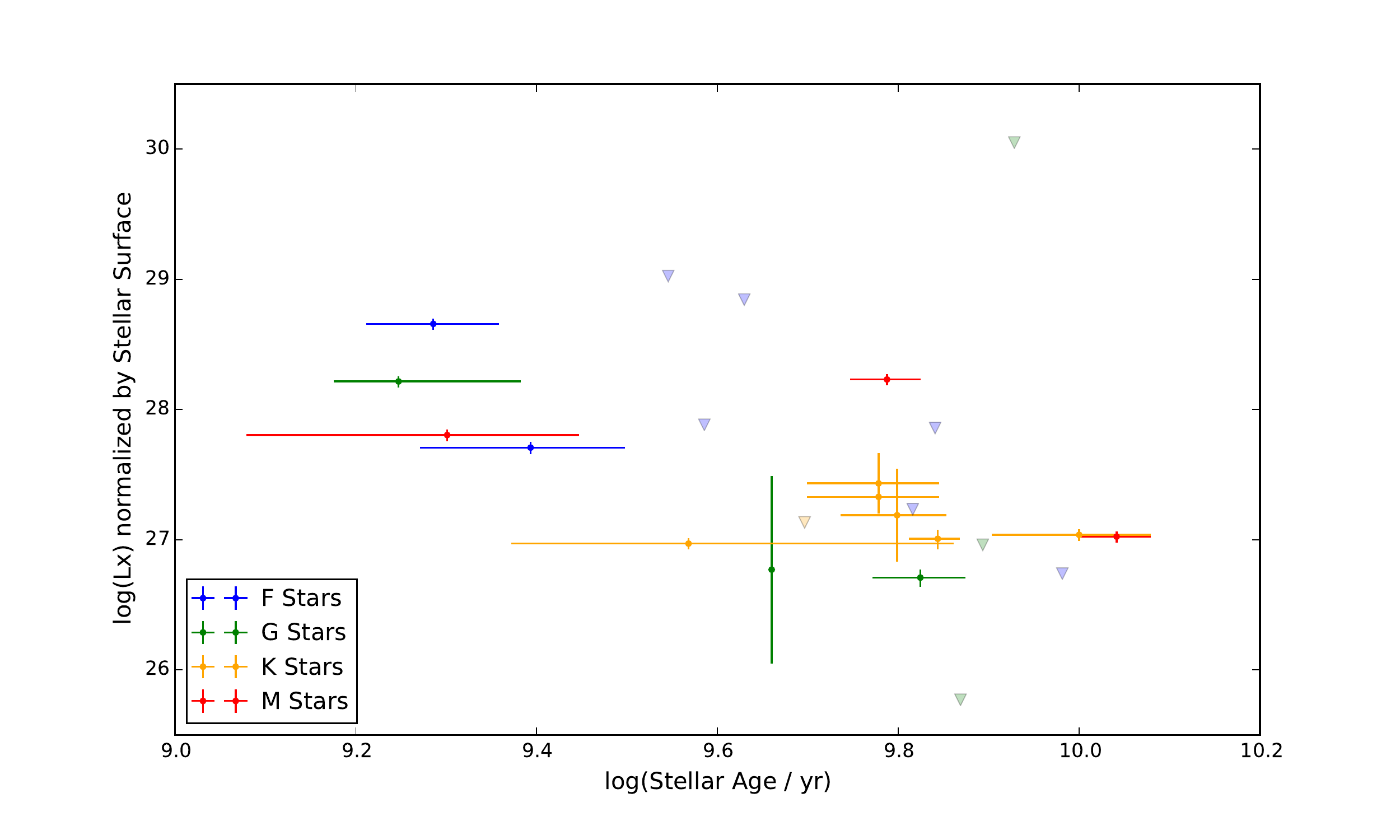}
\caption{Logarithmic plot of normalized X-ray Luminosity as a function of stellar age for the data analysed in this work; specifically, the quantity on the vertical axis is $\log\frac{L_{x}}{(R_\ast/R_\odot)^{2}}$. Upper limits are indicated by downward triangles and lighter colours.}
\centering
\label{data}
\end{figure*}

\begin{figure*}
\includegraphics[scale=0.65]{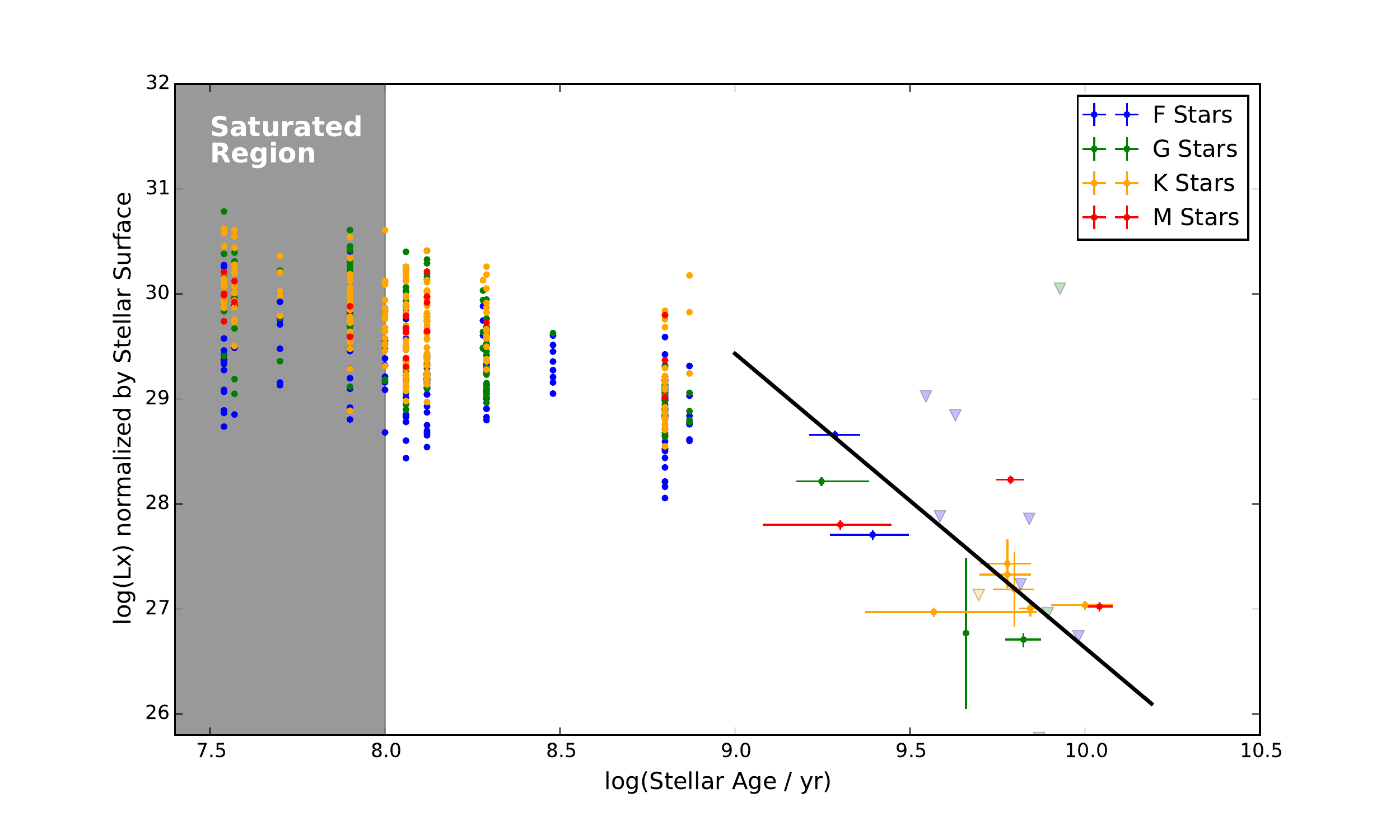}
\caption{Plot showing the data analysed in this work alongside the cluster data from \citet{jackson_lx_age}. The quantity on the vertical axis is $\log\frac{L_{x}}{(R_\ast/R_\odot)^{2}}$. Upper limits are indicated by downward triangles and lighter colours.  Black line indicates the best fit age-activity relationship found for data analysed in this work.}
\centering
\label{jack_plus}
\end{figure*}

\subsection{Fitting the Data}
In total, twenty four stars were fully analysed and are shown in Figure \ref{data}; the full details of the result for each star and stellar properties are listed in Appendix \ref{lx_results} and \ref{star_properties} respectively.

The majority of the sample stars have asteroseismic ages, only eight of the sample  have ages determined from other methods such as isochrone placement, white dwarf cooling times or association with a sub-population of stars in the galaxy. Ten stars in our sample have upper limits to their X-ray luminosities. The small number of X-ray detections demonstrates how difficult both the asteroseismic and X-ray measurements are. High-precision light-curves sampled at short cadences do not necessarily guarantee a precise measurement of age from asteroseismology. Even when a well-constrained age is determined, it does not guarantee an X-ray detection as resources are limited. Figure \ref{data} shows that the X-ray luminosity decreases with age as expected, but to gain more insight into what timescale this decrease occurs on the best fit relationship was found. 

Measurement errors are present in both the stellar age and the X-ray luminosity; we therefore performed an orthogonal distance regression (ODR) to fit the logarithmic X-ray luminosity (normalized by stellar surface) against the stellar age. In order to have normalized X-ray luminosities with values familiar to the reader, we chose to normalize by the stellar surface not in units of centimetres, but relative to the solar surface. This leads to normalized X-ray luminosities around $10^{27}$~erg\,s$^{-1}$\,$R_\odot^{-2}$. Only the 14 stars with X-ray detections or known X-ray luminosities were considered in this fit; for fitting purposes, we used symmetric errors in age and in $\log L_{\mathrm{X}}$. We obtained as the best fit relationship:

\begin{equation}
	\log\frac{L_{x}}{(R_\ast/R_\odot)^{2}} = 54.65 \pm 6.98 - (2.80 \pm 0.72)\log t
	\label{xray_age}
\end{equation}

We display this result visually in Figure \ref{jack_plus} where the fitted relationship is displayed as the solid black line. Upper limits are shown only for reference in the plot.

\section{Discussion}
\label{section5}

\citet{mamjek_08} (henceforth MH08) also derived an age-activity relationship for chromospheric activity derived from Ca~II H\&K emission. They also transform this into a relationship between X-ray luminosity and age, using a scaling relationship between chromospheric Ca~II H\&K and coronal X-ray emission, but no actual X-ray measurements of the sample stars are used. When we compare the $L_{x}$ ages derived from the age-activity relationship from MH08 to the literature ages for our sample, we find that the MH08 $L_{x}$ ages tend to be somewhat younger than the literature ages. Two factors may come into play here: the relationship between $R'_{HK}$ and X-ray luminosity in MH08 used very few stars which are less active than $\log R'_{HK} \approx -5$. Our sample contains very old stars which  we would expect to have chromospheric activity levels less than  $\log R'_{HK} \approx -5$, therefore probing a different part of the age/activity range than considered in the MH08 sample. Additionally, the use of their scaling relation between chromospheric and X-ray emission, which had been derived by \citet{1997AJ....114.1673S} for stars with activity levels $\log R'_{HK} > -5$ , may not fully catch the actual relation of those emissions for very old and inactive stars.

Figure \ref{jack_plus} shows the cluster data from \citet{jackson_lx_age} for ages below one gigayear, which we normalized by stellar surface area based on spectral type, alongside the sample of stars from this research for ages above one gigayear with our best fit age-activity relationship shown in black. As has been reported in many studies \citep{Vilhu1984, Jardine1999, Pizzolato:2003ga}, for very young stars there is a saturation of the X-ray luminosity until approximately 100Myr when the X-ray luminosity starts to decay. \citet{jackson_lx_age} quantitatively investigated the age-activity relationship using clusters as calibrators, normalizing by the bolometric luminosity and splitting into several mass bins. They found slope values for the age-activity relationship ranging from $-1.09 \pm 0.28$ to $-1.40 \pm 0.11$, considering seven spectral bins across the range from F-type stars to early M-type stars. Comparing these values to the slope value found in this work of $-2.80 \pm 0.72$, we find a steeper slope for the age-activity relationship at old ages than what is reported for any of the spectral bins for the younger stars. This steepening indicates a more rapid decay of stellar activity with age for cool stars older than a gigayear than for younger stars.

We will now discuss the implications of this steepening in relation to stellar spin-down and activity decrease in general. As mentioned previously the rotational velocity of a star will decrease over time as a result of magnetic braking where the rotation is related to the time (or age) by $v_{rot} \propto t^{-\alpha}$ where $\alpha = 0.5$ \citep{skumanich_72,Meibom:2011eu}. The first study of the relationship between rotation and activity was by \citealp{Pallavicini:1981hv} who found that $L_{x} \propto (vsin(i))^{1.9}$. Observations of solar-like stars confirmed that the relationship between activity and rotation takes the form of $L_{x} \propto v_{rot}^{\beta}$ where $\beta \approx 2$ \citep{Pizzolato:2003ga}. From these two relationships one can predict how X-ray luminosity varies with time as shown in Equation \eqref{theory_lx}.

\begin{equation}
	L_{x} \propto t^{-\alpha\beta}$  where $\alpha\beta \approx 1
	\label{theory_lx}
\end{equation}

Some previous studies have investigated the value of $\alpha\beta$. For example, \citealp{Guedel1997} studied nine solar-like G stars with ages ranging from 70 Myr to 9 Gyr (however they were constrained to rotation-inferred ages for most stars with ages beyond one gigayear) and found the value to be 1.5 for ages greater than 100 Myr. Later studies included \citealp{Giardino:2008jj} that studied the 1.5 Gyr NGC 752 cluster and presented results that were consistent with a value for $\alpha\beta$ of 1.5, but also found evidence for a steepening of the X-ray luminosity scaling law after the age of the Hyades cluster (625 million years). However, \citealp{Feigelson:2004kt} found an excellent fit for their data with a value for $\alpha\beta$ of 2 but also could not rule out the predicted value due to the small sample and systematic uncertainties.

The results from this research indicate that the value of $\alpha\beta$ for stars older than a gigayear is $2.80 \pm 0.72$, which is larger than the expected value of unity, and more in line with the direct investigations of $L_X$ versus age in the studies discussed in the previous paragraph. This leaves the challenge of explaining why the decay of magnetic activity is faster than predicted. One possibility is that the rotational spin down could be more rapid than expected from constant magnetised stellar winds \citep{Kawaler:1988fi}, i.e.\ $\alpha$ has a value greater than 0.5. \citet{Feigelson:2004kt} also postulated that the coronal mass ejections may contribute to stellar angular deceleration, changing the alpha exponent. But a recent study by \citet{2016Natur.529..181V} reports that there is weakened magnetic breaking for older late-type stars. Unfortunately, their sample and ours do not have sufficient overlap to compare age, rotation, and activity all together. A recent theoretical model by \citet{Blackman:2015bb} predicts weakened magnetic braking for older late-type stars; their model suggests that conductive losses are more important for these stars than wind losses which would imply a reduced angular momentum loss. Other theoretical work includes \citet{2015ApJ...813...40G} and \citet{2016MNRAS.455L..52V}, which show that the rotational spin down of a star may depend on the magnetic field geometry. If the rotational spin down was the only factor to affect the age-activity relationship then it should also show weakened magnetic braking and the exponent of the relationship should decrease and not increase as found in this work. From this evidence one would disfavour the more rapid rotational spin down as the cause of the more rapid decay of the magnetic activity.

Another possible explanation for the increased decay in magnetic activity for stars older than a gigayear is that the relationship between the X-ray luminosity and rotational velocity is not constant, i.e. the $\beta$ term changes as the star ages. There is some evidence for the steepening of the activity-rotation relationship, \citet{Wright:2011dj} considered a small, unbiased subset of their large sample of solar and late-type stars and found that a value for $\beta$ of 2.7 was a better fit for their data than the generally accepted value of 2. This was in agreement with \citet{Guedel1997} who found a value for $\beta$ of 2.64 for a sample of nine solar analogs and \citet{Feigelson:2004kt} who found an un-expectedly steep decay of X-ray emission as a function of age which could indicate a steepening of the activity-rotation relationship. However, more research is needed into the activity-rotation relationship to confirm if there is a steepening of the relationship as one of the lowest values considered in the \citet{Wright:2011dj} subset of data was of the Sun. In this research older stars have been considered that have lower X-ray luminosities than the Sun therefore the activity-rotation relationship needs to be extended to lower X-ray luminosities.

\section{Conclusions}
\label{section6}

In this work we have presented new X-ray detections of several old cool stars together with an analysis of archival data to form a sample of 24 cool stars with ages beyond one gigayear. Most stellar ages in our sample have been determined by asteroseismology, providing more accurate ages for old stars than most other studies were able to provide. We have investigated the age-activity relationship of these stars using observations from the \textit{Chandra} and \textit{XMM-Newton} X-ray telescopes. X-ray luminosities were determined for fourteen stars primarily, and spectral modelling was performed for eight of those stars; upper limits to the X-ray luminosity were determined for a further ten stars. We normalized the X-ray luminosity of the sample stars by the stellar surface, in order to  perform an analysis across varying spectral types. We find an age-exponent of $\alpha\beta = 2.80 \pm 0.72$, which represents a steepening of the age-activity relationship compared to what is seen for stars in clusters with ages below one gigayear.  A possible explanation for this steepening of the age-activity relationship is that rotational spin down is more rapid than previously thought. However, a recent observational study \citep{2016Natur.529..181V} indicates that there is weakened magnetic braking for older cool stars. If this is indeed true, our data presents evidence that there is a strong steepening of the rotation-activity relationship at old stellar ages instead of the age-rotation relationship itself. In either case, the data we have presented here demonstrates that the relationship between stellar age and activity steepens towards old stellar ages. Combined studies of age, rotation, and activity will be able to shed light on which components of the relationship are responsible for this.

\section*{Acknowledgements}
We thank the anonymous reviewer for their detailed and insightful comments, which added significantly to the clarity of this paper.
The scientific results reported in this article are based on observations made by the Chandra X-ray Observatory and by \textit{XMM-Newton}, an ESA science mission with instruments and contributions directly funded by ESA Member States and NASA. Support for this work was provided by the National Aeronautics and Space Administration through Chandra Award Number GO5-16101X issued by the \textit{Chandra} X-ray Observatory Center, which is operated by the Smithsonian Astrophysical Observatory for and on behalf of the National Aeronautics Space Administration under contract NAS8-03060.
This work has made use of data from the European Space Agency (ESA) mission {\it Gaia} (\url{http://www.cosmos.esa.int/gaia}), processed by the {\it Gaia} Data Processing and Analysis Consortium (DPAC, \url{http://www.cosmos.esa.int/web/gaia/dpac/consortium}). Funding for the DPAC has been provided by national institutions, in particular the institutions participating in the {\it Gaia} Multilateral Agreement.
This research made use of public databases hosted by SIMBAD, maintained by CDS, Strasbourg, France.
RB acknowledges funding from DE. CAW acknowledges support from the STFC grant ST/L000709/1. Funding for the Stellar Astrophysics Centre is provided by The Danish National Research Foundation (Grant agreement no. DNRF106). V.S.A.\ acknowledges support from Villum Fonden (research grant 10118).

\bibliographystyle{mnras}
\bibliography{references}

\begin{thebibliography}{}
\makeatletter
\relax
\def\mn@urlcharsother{\let\do\@makeother \do\$\do\&\do\#\do\^\do\_\do\%\do\~}
\def\mn@doi{\begingroup\mn@urlcharsother \@ifnextchar [ {\mn@doi@}
  {\mn@doi@[]}}
\def\mn@doi@[#1]#2{\def\@tempa{#1}\ifx\@tempa\@empty \href
  {http://dx.doi.org/#2} {doi:#2}\else \href {http://dx.doi.org/#2} {#1}\fi
  \endgroup}
\def\mn@eprint#1#2{\mn@eprint@#1:#2::\@nil}
\def\mn@eprint@arXiv#1{\href {http://arxiv.org/abs/#1} {{\tt arXiv:#1}}}
\def\mn@eprint@dblp#1{\href {http://dblp.uni-trier.de/rec/bibtex/#1.xml}
  {dblp:#1}}
\def\mn@eprint@#1:#2:#3:#4\@nil{\def\@tempa {#1}\def\@tempb {#2}\def\@tempc
  {#3}\ifx \@tempc \@empty \let \@tempc \@tempb \let \@tempb \@tempa \fi \ifx
  \@tempb \@empty \def\@tempb {arXiv}\fi \@ifundefined
  {mn@eprint@\@tempb}{\@tempb:\@tempc}{\expandafter \expandafter \csname
  mn@eprint@\@tempb\endcsname \expandafter{\@tempc}}}

\bibitem[\protect\citeauthoryear{{Anglada-Escud{\'e}}
  et~al.,}{{Anglada-Escud{\'e}} et~al.}{2014}]{gj_191_age}
{Anglada-Escud{\'e}} G.,  et~al., 2014, \mn@doi [MNRAS]
  {10.1093/mnrasl/slu076}, \href
  {http://adsabs.harvard.edu/abs/2014MNRAS.443L..89A} {443, L89}

\bibitem[\protect\citeauthoryear{{Angus}, {Aigrain}, {Foreman-Mackey}  \&
  {McQuillan}}{{Angus} et~al.}{2015}]{2015MNRAS.450.1787A}
{Angus} R.,  {Aigrain} S.,  {Foreman-Mackey} D.,   {McQuillan} A.,  2015,
  \mn@doi [MNRAS] {10.1093/mnras/stv423}, \href
  {http://adsabs.harvard.edu/abs/2015MNRAS.450.1787A} {450, 1787}

\bibitem[\protect\citeauthoryear{{Bahcall}, {Pinsonneault}  \&
  {Wasserburg}}{{Bahcall} et~al.}{1995}]{1995RvMP...67..781B}
{Bahcall} J.~N.,  {Pinsonneault} M.~H.,   {Wasserburg} G.~J.,  1995, \mn@doi
  [Reviews of Modern Physics] {10.1103/RevModPhys.67.781}, \href
  {http://adsabs.harvard.edu/abs/1995RvMP...67..781B} {67, 781}

\bibitem[\protect\citeauthoryear{{Barnes}}{{Barnes}}{2003}]{barnes_03}
{Barnes} S.~A.,  2003, \mn@doi [ApJ] {10.1086/367639}, \href
  {http://adsabs.harvard.edu/abs/2003ApJ...586..464B} {586, 464}

\bibitem[\protect\citeauthoryear{{Barnes}}{{Barnes}}{2007}]{2007ApJ...669.1167B}
{Barnes} S.~A.,  2007, \mn@doi [ApJ] {10.1086/519295}, \href
  {http://adsabs.harvard.edu/abs/2007ApJ...669.1167B} {669, 1167}

\bibitem[\protect\citeauthoryear{{Barnes} \& {Kim}}{{Barnes} \&
  {Kim}}{2010}]{Barnes:2010bf}
{Barnes} S.~A.,  {Kim} Y.-C.,  2010, \mn@doi [ApJ]
  {10.1088/0004-637X/721/1/675}, \href
  {http://adsabs.harvard.edu/abs/2010ApJ...721..675B} {721, 675}

\bibitem[\protect\citeauthoryear{{Barnes}, {Weingrill}, {Fritzewski},
  {Strassmeier}  \& {Platais}}{{Barnes} et~al.}{2016}]{2016ApJ...823...16B}
{Barnes} S.~A.,  {Weingrill} J.,  {Fritzewski} D.,  {Strassmeier} K.~G.,
  {Platais} I.,  2016, \mn@doi [ApJ] {10.3847/0004-637X/823/1/16}, \href
  {http://adsabs.harvard.edu/abs/2016ApJ...823...16B} {823, 16}

\bibitem[\protect\citeauthoryear{{Bergeron} et~al.,}{{Bergeron}
  et~al.}{2011}]{Bergeron2011}
{Bergeron} P.,  et~al., 2011, \mn@doi [ApJ] {10.1088/0004-637X/737/1/28}, \href
  {http://adsabs.harvard.edu/abs/2011ApJ...737...28B} {737, 28}

\bibitem[\protect\citeauthoryear{{Bertelli}, {Girardi}, {Marigo}  \&
  {Nasi}}{{Bertelli} et~al.}{2008}]{Bertelli2008}
{Bertelli} G.,  {Girardi} L.,  {Marigo} P.,   {Nasi} E.,  2008, \mn@doi
  [A{\&}A] {10.1051/0004-6361:20079165}, \href
  {http://adsabs.harvard.edu/abs/2008A%26A...484..815B} {484, 815}

\bibitem[\protect\citeauthoryear{{Blackman} \& {Owen}}{{Blackman} \&
  {Owen}}{2016}]{Blackman:2015bb}
{Blackman} E.~G.,  {Owen} J.~E.,  2016, \mn@doi [MNRAS] {10.1093/mnras/stw369},
  \href {http://adsabs.harvard.edu/abs/2016MNRAS.458.1548B} {458, 1548}

\bibitem[\protect\citeauthoryear{{Bouvier}, {Forestini}  \& {Allain}}{{Bouvier}
  et~al.}{1997}]{Bouvier1997}
{Bouvier} J.,  {Forestini} M.,   {Allain} S.,  1997, A{\&}A, \href
  {http://adsabs.harvard.edu/abs/1997A%26A...326.1023B} {326, 1023}

\bibitem[\protect\citeauthoryear{{Buchhave} \& {Latham}}{{Buchhave} \&
  {Latham}}{2015}]{2015ApJ...808..187B}
{Buchhave} L.~A.,  {Latham} D.~W.,  2015, \mn@doi [ApJ]
  {10.1088/0004-637X/808/2/187}, \href
  {http://adsabs.harvard.edu/abs/2015ApJ...808..187B} {808, 187}

\bibitem[\protect\citeauthoryear{{Chaplin} \& {Miglio}}{{Chaplin} \&
  {Miglio}}{2013}]{chaplin_astero}
{Chaplin} W.~J.,  {Miglio} A.,  2013, \mn@doi [ARA{\&}A]
  {10.1146/annurev-astro-082812-140938}, \href
  {http://adsabs.harvard.edu/abs/2013ARA%26A..51..353C} {51, 353}

\bibitem[\protect\citeauthoryear{{Chaplin} et~al.,}{{Chaplin}
  et~al.}{2014}]{astero2}
{Chaplin} W.~J.,  et~al., 2014, \mn@doi [\apjs] {10.1088/0067-0049/210/1/1},
  \href {http://adsabs.harvard.edu/abs/2014ApJS..210....1C} {210, 1}

\bibitem[\protect\citeauthoryear{{Cuntz}, {Saar}  \& {Musielak}}{{Cuntz}
  et~al.}{2000}]{Cuntz2000}
{Cuntz} M.,  {Saar} S.~H.,   {Musielak} Z.~E.,  2000, \mn@doi [ApJ]
  {10.1086/312609}, \href {http://adsabs.harvard.edu/abs/2000ApJ...533L.151C}
  {533, L151}

\bibitem[\protect\citeauthoryear{{DeWarf}, {Datin}  \& {Guinan}}{{DeWarf}
  et~al.}{2010}]{2010_dewarf}
{DeWarf} L.~E.,  {Datin} K.~M.,   {Guinan} E.~F.,  2010, \mn@doi [ApJ]
  {10.1088/0004-637X/722/1/343}, \href
  {http://adsabs.harvard.edu/abs/2010ApJ...722..343D} {722, 343}

\bibitem[\protect\citeauthoryear{{Erd{\'e}lyi} \& {Ballai}}{{Erd{\'e}lyi} \&
  {Ballai}}{2007}]{Erdelyi:2007gw}
{Erd{\'e}lyi} R.,  {Ballai} I.,  2007, \mn@doi [Astronomische Nachrichten]
  {10.1002/asna.200710803}, \href
  {http://adsabs.harvard.edu/abs/2007AN....328..726E} {328, 726}

\bibitem[\protect\citeauthoryear{{Feigelson} et~al.,}{{Feigelson}
  et~al.}{2004}]{Feigelson:2004kt}
{Feigelson} E.~D.,  et~al., 2004, \mn@doi [ApJ] {10.1086/422248}, \href
  {http://adsabs.harvard.edu/abs/2004ApJ...611.1107F} {611, 1107}

\bibitem[\protect\citeauthoryear{{Feltzing} \& {Holmberg}}{{Feltzing} \&
  {Holmberg}}{2000}]{gj_176_age}
{Feltzing} S.,  {Holmberg} J.,  2000, A{\&}A, \href
  {http://adsabs.harvard.edu/abs/2000A%26A...357..153F} {357, 153}

\bibitem[\protect\citeauthoryear{{Fouque} \& {Gieren}}{{Fouque} \&
  {Gieren}}{1997}]{1997A&A...320..799F}
{Fouque} P.,  {Gieren} W.~P.,  1997, A{\&}A, \href
  {http://adsabs.harvard.edu/abs/1997A%26A...320..799F} {320, 799}

\bibitem[\protect\citeauthoryear{{Fruscione} et~al.,}{{Fruscione}
  et~al.}{2006}]{ciao}
{Fruscione} A.,  et~al., 2006, in Society of Photo-Optical Instrumentation
  Engineers (SPIE) Conference Series. , \mn@doi{10.1117/12.671760}

\bibitem[\protect\citeauthoryear{{Fuhrmeister}, {Lalitha}, {Poppenhaeger},
  {Rudolf}, {Liefke}, {Reiners}, {Schmitt}  \& {Ness}}{{Fuhrmeister}
  et~al.}{2011}]{2011A&A...534A.133F}
{Fuhrmeister} B.,  {Lalitha} S.,  {Poppenhaeger} K.,  {Rudolf} N.,  {Liefke}
  C.,  {Reiners} A.,  {Schmitt} J.~H.~M.~M.,   {Ness} J.-U.,  2011, \mn@doi
  [A{\&}A] {10.1051/0004-6361/201117447}, \href
  {http://adsabs.harvard.edu/abs/2011A%26A...534A.133F} {534, A133}

\bibitem[\protect\citeauthoryear{{Gaia Collaboration} et~al.,}{{Gaia
  Collaboration} et~al.}{2016}]{2016A&A...595A...2G}
{Gaia Collaboration} et~al., 2016, \mn@doi [A{\&}A]
  {10.1051/0004-6361/201629512}, \href
  {http://adsabs.harvard.edu/abs/2016A%26A...595A...2G} {595, A2}

\bibitem[\protect\citeauthoryear{{Garc{\'e}s}, {Catal{\'a}n}  \&
  {Ribas}}{{Garc{\'e}s} et~al.}{2011}]{garces_wd}
{Garc{\'e}s} A.,  {Catal{\'a}n} S.,   {Ribas} I.,  2011, \mn@doi [A{\&}A]
  {10.1051/0004-6361/201116775}, \href
  {http://adsabs.harvard.edu/abs/2011A%26A...531A...7G} {531, A7}

\bibitem[\protect\citeauthoryear{{Garraffo}, {Drake}  \& {Cohen}}{{Garraffo}
  et~al.}{2015}]{2015ApJ...813...40G}
{Garraffo} C.,  {Drake} J.~J.,   {Cohen} O.,  2015, \mn@doi [ApJ]
  {10.1088/0004-637X/813/1/40}, \href
  {http://adsabs.harvard.edu/abs/2015ApJ...813...40G} {813, 40}

\bibitem[\protect\citeauthoryear{{Giardino}, {Pillitteri}, {Favata}  \&
  {Micela}}{{Giardino} et~al.}{2008}]{Giardino:2008jj}
{Giardino} G.,  {Pillitteri} I.,  {Favata} F.,   {Micela} G.,  2008, \mn@doi
  [A{\&}A] {10.1051/0004-6361:200810042}, \href
  {http://adsabs.harvard.edu/abs/2008A%26A...490..113G} {490, 113}

\bibitem[\protect\citeauthoryear{Gray}{Gray}{2005}]{CBO9781316036570A207}
Gray D.~F.,  2005, in , The Observation and Analysis of Stellar Photospheres,
  third edn, Cambridge University Press, pp 365--383

\bibitem[\protect\citeauthoryear{{Grevesse} \& {Sauval}}{{Grevesse} \&
  {Sauval}}{1998}]{1998SSRv...85..161G}
{Grevesse} N.,  {Sauval} A.~J.,  1998, \mn@doi [SSR] {10.1023/A:1005161325181},
  \href {http://adsabs.harvard.edu/abs/1998SSRv...85..161G} {85, 161}

\bibitem[\protect\citeauthoryear{{G{\"u}del}, {Guinan}  \&
  {Skinner}}{{G{\"u}del} et~al.}{1997}]{Guedel1997}
{G{\"u}del} M.,  {Guinan} E.~F.,   {Skinner} S.~L.,  1997, ApJ, \href
  {http://adsabs.harvard.edu/abs/1997ApJ...483..947G} {483, 947}

\bibitem[\protect\citeauthoryear{{Guinan}, {Engle}  \& {Durbin}}{{Guinan}
  et~al.}{2016}]{Guinan2016}
{Guinan} E.~F.,  {Engle} S.~G.,   {Durbin} A.,  2016, \mn@doi [\apj]
  {10.3847/0004-637X/821/2/81}, \href
  {http://adsabs.harvard.edu/abs/2016ApJ...821...81G} {821, 81}

\bibitem[\protect\citeauthoryear{{Herbst}, {Eisl{\"o}ffel}, {Mundt}  \&
  {Scholz}}{{Herbst} et~al.}{2007}]{Herbst2007}
{Herbst} W.,  {Eisl{\"o}ffel} J.,  {Mundt} R.,   {Scholz} A.,  2007, Protostars
  and Planets V, \href {http://adsabs.harvard.edu/abs/2007prpl.conf..297H} {pp
  297--311}

\bibitem[\protect\citeauthoryear{{Holberg} \& {Bergeron}}{{Holberg} \&
  {Bergeron}}{2006}]{Holberg2006}
{Holberg} J.~B.,  {Bergeron} P.,  2006, \mn@doi [AJ] {10.1086/505938}, \href
  {http://adsabs.harvard.edu/abs/2006AJ....132.1221H} {132, 1221}

\bibitem[\protect\citeauthoryear{{Jackson}, {Davis}  \& {Wheatley}}{{Jackson}
  et~al.}{2012}]{jackson_lx_age}
{Jackson} A.~P.,  {Davis} T.~A.,   {Wheatley} P.~J.,  2012, \mn@doi [MNRAS]
  {10.1111/j.1365-2966.2012.20657.x}, \href
  {http://adsabs.harvard.edu/abs/2012MNRAS.422.2024J} {422, 2024}

\bibitem[\protect\citeauthoryear{{Jansen} et~al.,}{{Jansen}
  et~al.}{2001}]{Jansen2001}
{Jansen} F.,  et~al., 2001, \mn@doi [A{\&}A] {10.1051/0004-6361:20000036},
  \href {http://adsabs.harvard.edu/abs/2001A%26A...365L...1J} {365, L1}

\bibitem[\protect\citeauthoryear{{Jardine} \& {Unruh}}{{Jardine} \&
  {Unruh}}{1999}]{Jardine1999}
{Jardine} M.,  {Unruh} Y.~C.,  1999, A{\&}A, \href
  {http://adsabs.harvard.edu/abs/1999A%26A...346..883J} {346, 883}

\bibitem[\protect\citeauthoryear{{Johnstone} \& {G{\"u}del}}{{Johnstone} \&
  {G{\"u}del}}{2015}]{Johnstone2015}
{Johnstone} C.~P.,  {G{\"u}del} M.,  2015, \mn@doi [A{\&}A]
  {10.1051/0004-6361/201425283}, \href
  {http://adsabs.harvard.edu/abs/2015A%26A...578A.129J} {578, A129}

\bibitem[\protect\citeauthoryear{{Kalirai}, {Hansen}, {Kelson}, {Reitzel},
  {Rich}  \& {Richer}}{{Kalirai} et~al.}{2008}]{2008ApJ...676..594K}
{Kalirai} J.~S.,  {Hansen} B.~M.~S.,  {Kelson} D.~D.,  {Reitzel} D.~B.,  {Rich}
  R.~M.,   {Richer} H.~B.,  2008, \mn@doi [\apj] {10.1086/527028}, \href
  {http://adsabs.harvard.edu/abs/2008ApJ...676..594K} {676, 594}

\bibitem[\protect\citeauthoryear{{Kawaler}}{{Kawaler}}{1988}]{Kawaler:1988fi}
{Kawaler} S.~D.,  1988, \mn@doi [ApJ] {10.1086/166740}, \href
  {http://adsabs.harvard.edu/abs/1988ApJ...333..236K} {333, 236}

\bibitem[\protect\citeauthoryear{{Kervella} et~al.,}{{Kervella}
  et~al.}{2008}]{61_cyg_age}
{Kervella} P.,  et~al., 2008, \mn@doi [A{\&}A] {10.1051/0004-6361:200810080},
  \href {http://adsabs.harvard.edu/abs/2008A%26A...488..667K} {488, 667}

\bibitem[\protect\citeauthoryear{{Kowalski} \& {Saumon}}{{Kowalski} \&
  {Saumon}}{2006}]{Kowalski2006}
{Kowalski} P.~M.,  {Saumon} D.,  2006, \mn@doi [ApJ] {10.1086/509723}, \href
  {http://adsabs.harvard.edu/abs/2006ApJ...651L.137K} {651, L137}

\bibitem[\protect\citeauthoryear{{Lorenzo-Oliveira}, {Porto de Mello}  \&
  {Schiavon}}{{Lorenzo-Oliveira} et~al.}{2016}]{2016A&A...594L...3L}
{Lorenzo-Oliveira} D.,  {Porto de Mello} G.~F.,   {Schiavon} R.~P.,  2016,
  \mn@doi [A{\&}A] {10.1051/0004-6361/201629233}, \href
  {http://adsabs.harvard.edu/abs/2016A%26A...594L...3L} {594, L3}

\bibitem[\protect\citeauthoryear{{Mamajek} \& {Hillenbrand}}{{Mamajek} \&
  {Hillenbrand}}{2008}]{mamjek_08}
{Mamajek} E.~E.,  {Hillenbrand} L.~A.,  2008, \mn@doi [ApJ] {10.1086/591785},
  \href {http://adsabs.harvard.edu/abs/2008ApJ...687.1264M} {687, 1264}

\bibitem[\protect\citeauthoryear{{Mathur} et~al.,}{{Mathur}
  et~al.}{2012}]{Mathur:2012bj}
{Mathur} S.,  et~al., 2012, \mn@doi [ApJ] {10.1088/0004-637X/749/2/152}, \href
  {http://adsabs.harvard.edu/abs/2012ApJ...749..152M} {749, 152}

\bibitem[\protect\citeauthoryear{{Matt}, {Brun}, {Baraffe}, {Bouvier}  \&
  {Chabrier}}{{Matt} et~al.}{2015}]{2015ApJ...799L..23M}
{Matt} S.~P.,  {Brun} A.~S.,  {Baraffe} I.,  {Bouvier} J.,   {Chabrier} G.,
  2015, \mn@doi [ApJ] {10.1088/2041-8205/799/2/L23}, \href
  {http://adsabs.harvard.edu/abs/2015ApJ...799L..23M} {799, L23}

\bibitem[\protect\citeauthoryear{{Meibom} et~al.,}{{Meibom}
  et~al.}{2011}]{Meibom:2011eu}
{Meibom} S.,  et~al., 2011, \mn@doi [ApJ] {10.1088/2041-8205/733/1/L9}, \href
  {http://adsabs.harvard.edu/abs/2011ApJ...733L...9M} {733, L9}

\bibitem[\protect\citeauthoryear{{Meibom}, {Barnes}, {Platais}, {Gilliland},
  {Latham}  \& {Mathieu}}{{Meibom} et~al.}{2015}]{Meibom:2015if}
{Meibom} S.,  {Barnes} S.~A.,  {Platais} I.,  {Gilliland} R.~L.,  {Latham}
  D.~W.,   {Mathieu} R.~D.,  2015, \mn@doi [Nature] {10.1038/nature14118},
  \href {http://adsabs.harvard.edu/abs/2015Natur.517..589M} {517, 589}

\bibitem[\protect\citeauthoryear{{Metcalfe} et~al.,}{{Metcalfe}
  et~al.}{2012}]{Metcalfe:2012kv}
{Metcalfe} T.~S.,  et~al., 2012, \mn@doi [ApJ] {10.1088/2041-8205/748/1/L10},
  \href {http://adsabs.harvard.edu/abs/2012ApJ...748L..10M} {748, L10}

\bibitem[\protect\citeauthoryear{{Miglio} \& {Montalb{\'a}n}}{{Miglio} \&
  {Montalb{\'a}n}}{2005}]{prox_cen_age_2}
{Miglio} A.,  {Montalb{\'a}n} J.,  2005, \mn@doi [A{\&}A]
  {10.1051/0004-6361:20052988}, \href
  {http://adsabs.harvard.edu/abs/2005A%26A...441..615M} {441, 615}

\bibitem[\protect\citeauthoryear{{Newton}, {Irwin}, {Charbonneau}, {Berlind},
  {Calkins}  \& {Mink}}{{Newton} et~al.}{2017}]{2017ApJ...834...85N}
{Newton} E.~R.,  {Irwin} J.,  {Charbonneau} D.,  {Berlind} P.,  {Calkins}
  M.~L.,   {Mink} J.,  2017, \mn@doi [ApJ] {10.3847/1538-4357/834/1/85}, \href
  {http://adsabs.harvard.edu/abs/2017ApJ...834...85N} {834, 85}

\bibitem[\protect\citeauthoryear{{Noyes}, {Hartmann}, {Baliunas}, {Duncan}  \&
  {Vaughan}}{{Noyes} et~al.}{1984}]{1984ApJ...279..763N}
{Noyes} R.~W.,  {Hartmann} L.~W.,  {Baliunas} S.~L.,  {Duncan} D.~K.,
  {Vaughan} A.~H.,  1984, \mn@doi [ApJ] {10.1086/161945}, \href
  {http://adsabs.harvard.edu/abs/1984ApJ...279..763N} {279, 763}

\bibitem[\protect\citeauthoryear{{Pace}}{{Pace}}{2013}]{pace_cahk_age}
{Pace} G.,  2013, \mn@doi [A{\&}A] {10.1051/0004-6361/201220364}, \href
  {http://adsabs.harvard.edu/abs/2013A%26A...551L...8P} {551, L8}

\bibitem[\protect\citeauthoryear{{Pallavicini}, {Golub}, {Rosner}, {Vaiana},
  {Ayres}  \& {Linsky}}{{Pallavicini} et~al.}{1981}]{Pallavicini:1981hv}
{Pallavicini} R.,  {Golub} L.,  {Rosner} R.,  {Vaiana} G.~S.,  {Ayres} T.,
  {Linsky} J.~L.,  1981, \mn@doi [ApJ] {10.1086/159152}, \href
  {http://adsabs.harvard.edu/abs/1981ApJ...248..279P} {248, 279}

\bibitem[\protect\citeauthoryear{{Parker}}{{Parker}}{1955}]{parker_55}
{Parker} E.~N.,  1955, \mn@doi [ApJ] {10.1086/146087}, \href
  {http://adsabs.harvard.edu/abs/1955ApJ...122..293P} {122, 293}

\bibitem[\protect\citeauthoryear{{Pecaut} \& {Mamajek}}{{Pecaut} \&
  {Mamajek}}{2013}]{2013ApJS..208....9P}
{Pecaut} M.~J.,  {Mamajek} E.~E.,  2013, \mn@doi [ApJS]
  {10.1088/0067-0049/208/1/9}, \href
  {http://adsabs.harvard.edu/abs/2013ApJS..208....9P} {208, 9}

\bibitem[\protect\citeauthoryear{{Peres}, {Orlando}, {Reale}, {Rosner}  \&
  {Hudson}}{{Peres} et~al.}{2000}]{2000ApJ...528..537P}
{Peres} G.,  {Orlando} S.,  {Reale} F.,  {Rosner} R.,   {Hudson} H.,  2000,
  \mn@doi [ApJ] {10.1086/308136}, \href
  {http://adsabs.harvard.edu/abs/2000ApJ...528..537P} {528, 537}

\bibitem[\protect\citeauthoryear{{Pillitteri}, {Maggio}, {Micela}, {Sciortino},
  {Wolk}  \& {Matsakos}}{{Pillitteri} et~al.}{2015}]{Pillitteri2015}
{Pillitteri} I.,  {Maggio} A.,  {Micela} G.,  {Sciortino} S.,  {Wolk} S.~J.,
  {Matsakos} T.,  2015, \mn@doi [ApJ] {10.1088/0004-637X/805/1/52}, \href
  {http://adsabs.harvard.edu/abs/2015ApJ...805...52P} {805, 52}

\bibitem[\protect\citeauthoryear{{Pizzolato}, {Maggio}, {Micela}, {Sciortino}
  \& {Ventura}}{{Pizzolato} et~al.}{2003}]{Pizzolato:2003ga}
{Pizzolato} N.,  {Maggio} A.,  {Micela} G.,  {Sciortino} S.,   {Ventura} P.,
  2003, \mn@doi [A{\&}A] {10.1051/0004-6361:20021560}, \href
  {http://adsabs.harvard.edu/abs/2003A%26A...397..147P} {397, 147}

\bibitem[\protect\citeauthoryear{{Poppenhaeger} \& {Wolk}}{{Poppenhaeger} \&
  {Wolk}}{2014}]{Poppenhaeger2014}
{Poppenhaeger} K.,  {Wolk} S.~J.,  2014, \mn@doi [A{\&}A]
  {10.1051/0004-6361/201423454}, \href
  {http://adsabs.harvard.edu/abs/2014A%26A...565L...1P} {565, L1}

\bibitem[\protect\citeauthoryear{{Poppenhaeger}, {Czesla}, {Schr{\"o}ter},
  {Lalitha}, {Kashyap}  \& {Schmitt}}{{Poppenhaeger}
  et~al.}{2012}]{2012A&A...541A..26P}
{Poppenhaeger} K.,  {Czesla} S.,  {Schr{\"o}ter} S.,  {Lalitha} S.,  {Kashyap}
  V.,   {Schmitt} J.~H.~M.~M.,  2012, \mn@doi [A{\&}A]
  {10.1051/0004-6361/201118507}, \href
  {http://adsabs.harvard.edu/abs/2012A%26A...541A..26P} {541, A26}

\bibitem[\protect\citeauthoryear{{Preibisch} \& {Feigelson}}{{Preibisch} \&
  {Feigelson}}{2005}]{Preibisch2005}
{Preibisch} T.,  {Feigelson} E.~D.,  2005, \mn@doi [ApJS] {10.1086/432094},
  \href {http://adsabs.harvard.edu/abs/2005ApJS..160..390P} {160, 390}

\bibitem[\protect\citeauthoryear{{Raymond} \& {Smith}}{{Raymond} \&
  {Smith}}{1977}]{Raymond1977}
{Raymond} J.~C.,  {Smith} B.~W.,  1977, \mn@doi [ApJS] {10.1086/190486}, \href
  {http://adsabs.harvard.edu/abs/1977ApJS...35..419R} {35, 419}

\bibitem[\protect\citeauthoryear{{Reiners}, {Joshi}  \& {Goldman}}{{Reiners}
  et~al.}{2012}]{2012AJ....143...93R}
{Reiners} A.,  {Joshi} N.,   {Goldman} B.,  2012, \mn@doi [\aj]
  {10.1088/0004-6256/143/4/93}, \href
  {http://adsabs.harvard.edu/abs/2012AJ....143...93R} {143, 93}

\bibitem[\protect\citeauthoryear{{Robrade} \& {Schmitt}}{{Robrade} \&
  {Schmitt}}{2016}]{2016arXiv161206570R}
{Robrade} J.,  {Schmitt} J.~H.~M.~M.,  2016, preprint, \href
  {http://adsabs.harvard.edu/abs/2016arXiv161206570R} {} (\mn@eprint {arXiv}
  {1612.06570})

\bibitem[\protect\citeauthoryear{{Robrade}, {Schmitt}  \& {Favata}}{{Robrade}
  et~al.}{2012}]{2012A&A...543A..84R}
{Robrade} J.,  {Schmitt} J.~H.~M.~M.,   {Favata} F.,  2012, \mn@doi [A{\&}A]
  {10.1051/0004-6361/201219046}, \href
  {http://adsabs.harvard.edu/abs/2012A%26A...543A..84R} {543, A84}

\bibitem[\protect\citeauthoryear{{Rocha-Pinto} \& {Maciel}}{{Rocha-Pinto} \&
  {Maciel}}{1998}]{1998MNRAS.298..332R}
{Rocha-Pinto} H.~J.,  {Maciel} W.~J.,  1998, \mn@doi [MNRAS]
  {10.1046/j.1365-8711.1998.01597.x}, \href
  {http://adsabs.harvard.edu/abs/1998MNRAS.298..332R} {298, 332}

\bibitem[\protect\citeauthoryear{{Schatzman}}{{Schatzman}}{1962}]{1962AnAp...25...18S}
{Schatzman} E.,  1962, Annales d'Astrophysique, \href
  {http://adsabs.harvard.edu/abs/1962AnAp...25...18S} {25, 18}

\bibitem[\protect\citeauthoryear{{Schmitt} \& {Liefke}}{{Schmitt} \&
  {Liefke}}{2004}]{2004A&A...417..651S}
{Schmitt} J.~H.~M.~M.,  {Liefke} C.,  2004, \mn@doi [A{\&}A]
  {10.1051/0004-6361:20030495}, \href
  {http://adsabs.harvard.edu/abs/2004A%26A...417..651S} {417, 651}

\bibitem[\protect\citeauthoryear{{Schrijver}}{{Schrijver}}{1987}]{Schrijver1987}
{Schrijver} C.~J.,  1987, A{\&}A, \href
  {http://adsabs.harvard.edu/abs/1987A%26A...172..111S} {172, 111}

\bibitem[\protect\citeauthoryear{Schrijver \& Zwaan}{Schrijver \&
  Zwaan}{2000}]{9780511546037}
Schrijver C.~J.,  Zwaan C.,  2000, Solar and Stellar Magnetic Activity.
Cambridge University Press

\bibitem[\protect\citeauthoryear{{Silva Aguirre} et~al.,}{{Silva Aguirre}
  et~al.}{2015}]{astero1}
{Silva Aguirre} V.,  et~al., 2015, \mn@doi [\mnras] {10.1093/mnras/stv1388},
  \href {http://adsabs.harvard.edu/abs/2015MNRAS.452.2127S} {452, 2127}

\bibitem[\protect\citeauthoryear{{Silva Aguirre} et~al.,}{{Silva Aguirre}
  et~al.}{2017}]{SilvaAguirre2017}
{Silva Aguirre} V.,  et~al., 2017, \mn@doi [ApJ] {10.3847/1538-4357/835/2/173},
  \href {http://adsabs.harvard.edu/abs/2017ApJ...835..173S} {835, 173}

\bibitem[\protect\citeauthoryear{{Skumanich}}{{Skumanich}}{1972}]{skumanich_72}
{Skumanich} A.,  1972, \mn@doi [ApJ] {10.1086/151310}, \href
  {http://adsabs.harvard.edu/abs/1972ApJ...171..565S} {171, 565}

\bibitem[\protect\citeauthoryear{{Stelzer} \& {Neuh{\"a}user}}{{Stelzer} \&
  {Neuh{\"a}user}}{2001}]{2001A&A...377..538S}
{Stelzer} B.,  {Neuh{\"a}user} R.,  2001, \mn@doi [A{\&}A]
  {10.1051/0004-6361:20011093}, \href
  {http://adsabs.harvard.edu/abs/2001A%26A...377..538S} {377, 538}

\bibitem[\protect\citeauthoryear{{Sterzik} \& {Schmitt}}{{Sterzik} \&
  {Schmitt}}{1997}]{1997AJ....114.1673S}
{Sterzik} M.~F.,  {Schmitt} J.~H.~M.~M.,  1997, \mn@doi [\aj] {10.1086/118597},
  \href {http://adsabs.harvard.edu/abs/1997AJ....114.1673S} {114, 1673}

\bibitem[\protect\citeauthoryear{{Telleschi}, {G{\"u}del}, {Briggs}, {Audard},
  {Ness}  \& {Skinner}}{{Telleschi} et~al.}{2005}]{2005ApJ...622..653T}
{Telleschi} A.,  {G{\"u}del} M.,  {Briggs} K.,  {Audard} M.,  {Ness} J.-U.,
  {Skinner} S.~L.,  2005, \mn@doi [ApJ] {10.1086/428109}, \href
  {http://adsabs.harvard.edu/abs/2005ApJ...622..653T} {622, 653}

\bibitem[\protect\citeauthoryear{{Tremblay}, {Bergeron}  \&
  {Gianninas}}{{Tremblay} et~al.}{2011}]{Tremblay2011}
{Tremblay} P.-E.,  {Bergeron} P.,   {Gianninas} A.,  2011, \mn@doi [ApJ]
  {10.1088/0004-637X/730/2/128}, \href
  {http://adsabs.harvard.edu/abs/2011ApJ...730..128T} {730, 128}

\bibitem[\protect\citeauthoryear{{Tremblay}, {Ludwig}, {Steffen}  \&
  {Freytag}}{{Tremblay} et~al.}{2013}]{Tremblay2013}
{Tremblay} P.-E.,  {Ludwig} H.-G.,  {Steffen} M.,   {Freytag} B.,  2013,
  \mn@doi [A{\&}A] {10.1051/0004-6361/201322318}, \href
  {http://adsabs.harvard.edu/abs/2013A%26A...559A.104T} {559, A104}

\bibitem[\protect\citeauthoryear{{Vidotto} et~al.,}{{Vidotto}
  et~al.}{2016}]{2016MNRAS.455L..52V}
{Vidotto} A.~A.,  et~al., 2016, \mn@doi [MNRAS] {10.1093/mnrasl/slv147}, \href
  {http://adsabs.harvard.edu/abs/2016MNRAS.455L..52V} {455, L52}

\bibitem[\protect\citeauthoryear{{Vilhu}}{{Vilhu}}{1984}]{Vilhu1984}
{Vilhu} O.,  1984, A{\&}A, \href
  {http://adsabs.harvard.edu/abs/1984A%26A...133..117V} {133, 117}

\bibitem[\protect\citeauthoryear{{Watson}, {Littlefair}, {Collier Cameron},
  {Dhillon}  \& {Simpson}}{{Watson} et~al.}{2010}]{2010MNRAS.408.1606W}
{Watson} C.~A.,  {Littlefair} S.~P.,  {Collier Cameron} A.,  {Dhillon} V.~S.,
  {Simpson} E.~K.,  2010, \mn@doi [MNRAS] {10.1111/j.1365-2966.2010.17233.x},
  \href {http://adsabs.harvard.edu/abs/2010MNRAS.408.1606W} {408, 1606}

\bibitem[\protect\citeauthoryear{{Weisskopf}, {Tananbaum}, {Van Speybroeck}  \&
  {O'Dell}}{{Weisskopf} et~al.}{2000}]{Weisskopf2000}
{Weisskopf} M.~C.,  {Tananbaum} H.~D.,  {Van Speybroeck} L.~P.,   {O'Dell}
  S.~L.,  2000, in {J.~E.~Truemper \& B.~Aschenbach} ed.,  Society of
  Photo-Optical Instrumentation Engineers (SPIE) Conference Series Vol. 4012,
  Society of Photo-Optical Instrumentation Engineers (SPIE) Conference Series.
  pp 2--16 (\mn@eprint {} {arXiv:astro-ph/0004127})

\bibitem[\protect\citeauthoryear{{Wenger} et~al.,}{{Wenger}
  et~al.}{2000}]{2000A&AS..143....9W}
{Wenger} M.,  et~al., 2000, \mn@doi [\aaps] {10.1051/aas:2000332}, \href
  {http://adsabs.harvard.edu/abs/2000A%26AS..143....9W} {143, 9}

\bibitem[\protect\citeauthoryear{{Wright} \& {Drake}}{{Wright} \&
  {Drake}}{2016}]{2016Natur.535..526W}
{Wright} N.~J.,  {Drake} J.~J.,  2016, \mn@doi [Nature] {10.1038/nature18638},
  \href {http://adsabs.harvard.edu/abs/2016Natur.535..526W} {535, 526}

\bibitem[\protect\citeauthoryear{{Wright}, {Drake}, {Mamajek}  \&
  {Henry}}{{Wright} et~al.}{2011}]{Wright:2011dj}
{Wright} N.~J.,  {Drake} J.~J.,  {Mamajek} E.~E.,   {Henry} G.~W.,  2011,
  \mn@doi [ApJ] {10.1088/0004-637X/743/1/48}, \href
  {http://adsabs.harvard.edu/abs/2011ApJ...743...48W} {743, 48}

\bibitem[\protect\citeauthoryear{{Yelle}}{{Yelle}}{2004}]{2004Icar..170..167Y}
{Yelle} R.~V.,  2004, \mn@doi [Icarus] {10.1016/j.icarus.2004.02.008}, \href
  {http://adsabs.harvard.edu/abs/2004Icar..170..167Y} {170, 167}

\bibitem[\protect\citeauthoryear{{Zhao}, {Oswalt}, {Willson}, {Wang}  \&
  {Zhao}}{{Zhao} et~al.}{2012}]{2012ApJ...746..144Z}
{Zhao} J.~K.,  {Oswalt} T.~D.,  {Willson} L.~A.,  {Wang} Q.,   {Zhao} G.,
  2012, \mn@doi [ApJ] {10.1088/0004-637X/746/2/144}, \href
  {http://adsabs.harvard.edu/abs/2012ApJ...746..144Z} {746, 144}

\bibitem[\protect\citeauthoryear{{van Saders} \& {Pinsonneault}}{{van Saders}
  \& {Pinsonneault}}{2013}]{2013ApJ...776...67V}
{van Saders} J.~L.,  {Pinsonneault} M.~H.,  2013, \mn@doi [ApJ]
  {10.1088/0004-637X/776/2/67}, \href
  {http://adsabs.harvard.edu/abs/2013ApJ...776...67V} {776, 67}

\bibitem[\protect\citeauthoryear{{van Saders}, {Ceillier}, {Metcalfe}, {Silva
  Aguirre}, {Pinsonneault}, {Garc{\'{\i}}a}, {Mathur}  \& {Davies}}{{van
  Saders} et~al.}{2016}]{2016Natur.529..181V}
{van Saders} J.~L.,  {Ceillier} T.,  {Metcalfe} T.~S.,  {Silva Aguirre} V.,
  {Pinsonneault} M.~H.,  {Garc{\'{\i}}a} R.~A.,  {Mathur} S.,   {Davies} G.~R.,
   2016, \mn@doi [Nature] {10.1038/nature16168}, \href
  {http://adsabs.harvard.edu/abs/2016Natur.529..181V} {529, 181}

\makeatother
\end{thebibliography}

\bsp	

\clearpage

\begin{appendices}
\section{Additional Information on individual analysis of Stars}
\label{additional_info}

Section \ref{section3} details the methodology used to determine the X-ray luminosities for our sample of stars. Since the sample is made up of a number of different observations from both the \textit{XMM-Newton} and \textit{Chandra} telescopes, we present additional details about the individual observations and the data reduction here.

\textit{16 Cyg A:} This star was detected in combining two \textit{Chandra} observations on a front-illuminated CCD, which provides energy sensitivity only above 0.6 keV. We extrapolated the flux for the full 0.2-2 keV band, assuming a coronal temperature of $\log T = 6.5$. The star 16 Cyg B was covered by the same observations, but is undetected and we report the corresponding upper limit. There is also an earlier, shorter \textit{XMM-Newton} observation covering both stars, but they were both undetected in that observation, which is consistent with the \textit{Chandra} data.

\textit{40 Eri A:} This star is in a wide binary system with the white dwarf 40 Eri B, with a projected distance of $83^{\prime\prime}$ \citep{2000A&AS..143....9W}, translating to a physical distance of ca.\ 400 AU. Using the techniques outlined in section~\ref{wd_ages}, we use its reported $T_{eff}$ and $\log g$ values \citep{2012ApJ...746..144Z} to calculate an age of $3.70^{+3.57}_{-1.34}$~Gyr for 40 Eri B, which we adopt as the age of the system. 40 Eri A was observed with a back-illuminated \textit{Chandra} CCD, which provides energy sensitivity above 0.245 keV. We performed a spectral fit and extrapolated the stellar X-ray flux for the full 0.2-2 keV energy range. There is also a third star, 40 Eri C, present in the system; however, it is close enough to the white dwarf (ca.\ 35 AU) that its rotation and activity properties may have been affected during the evolution of the white dwarf progenitor, which is why we do not include 40 Eri C in our analysis. 

\begin{figure*}
\begin{subfigure}{1.0\columnwidth}
	\centering
	\includegraphics[height = 0.2\paperheight,width=0.9\columnwidth]{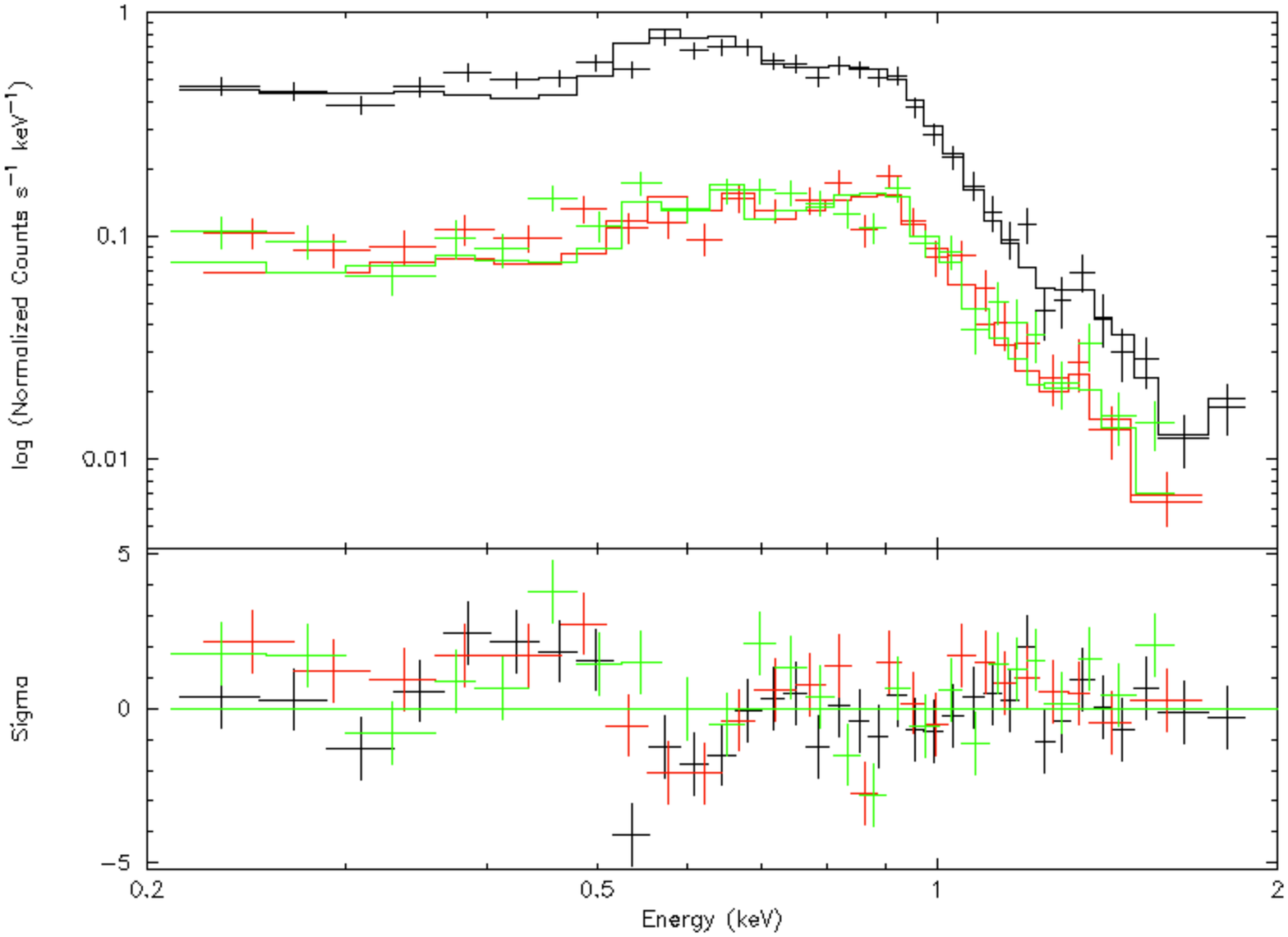}
	\caption{61 Cyg A}
\end{subfigure}
\begin{subfigure}{1.0\columnwidth}
\centering
\includegraphics[height = 0.2\paperheight,width=0.9\columnwidth]{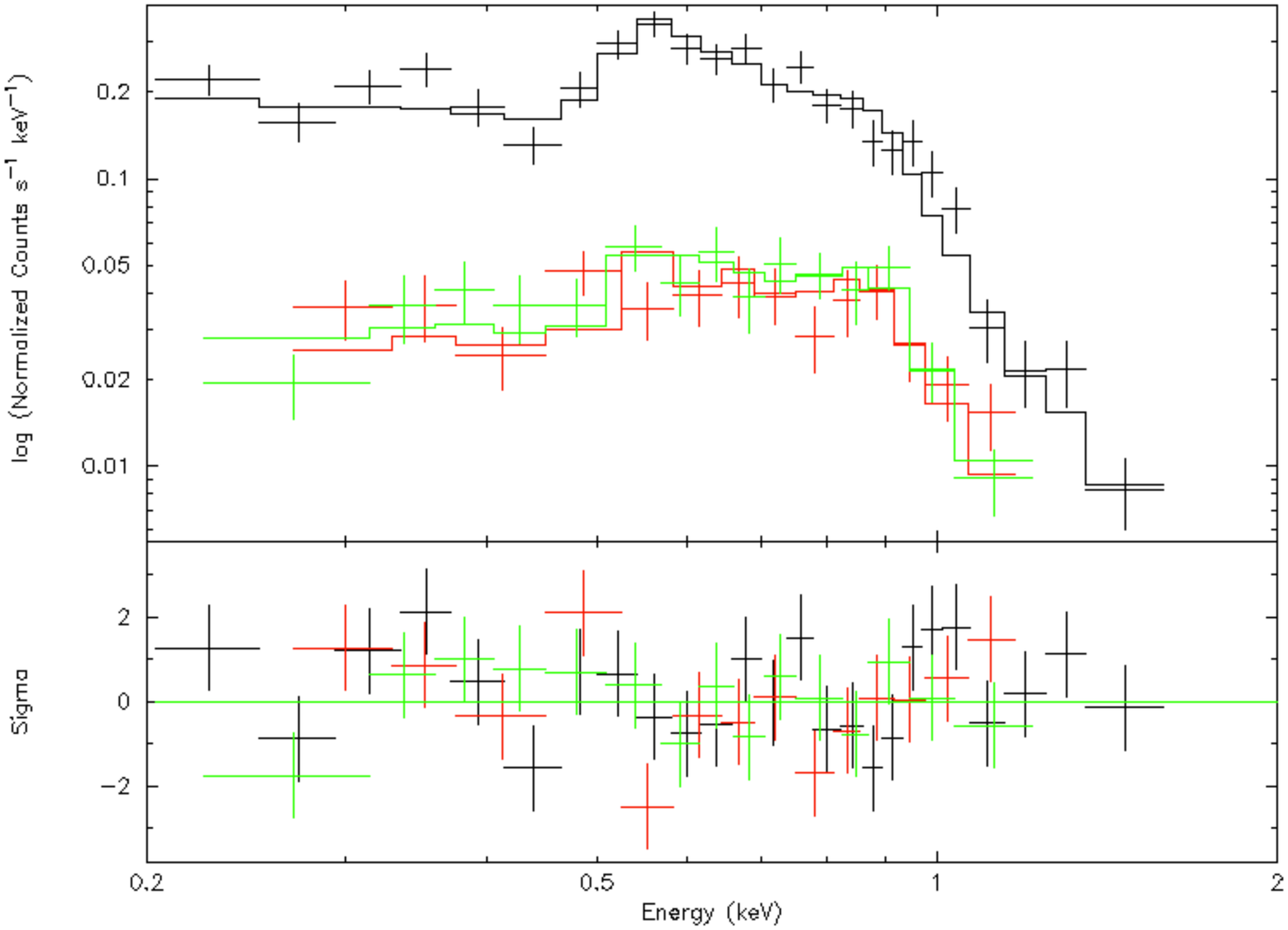}
\caption{61 Cyg B}
\end{subfigure}
\par\bigskip

\caption{X-ray spectra (grouped to 15 counts per bin) and best fit models for one exemplary observation of 61 Cyg A and B, respectively. These sources were detected to at least three sigma and contained over a hundred counts in the source region. The top region of each subplot shows the number of counts per second per keV as a function of energy. The bottom region of each subplot shows the sigma value for the best fit model as a function of energy. Different colours indicate spectra from different detectors which are fitted simultaneously to ensure a more accurate fit.\label{61cygfig}}

\label{lastpage}

\end{figure*}

\textit{61 Cyg A and B:} These stars have been monitored in X-rays with \textit{XMM-Newton} over several years. We show one exemplary X-ray spectrum for each star in Fig.~\ref{61cygfig}. 61 Cyg A has been found to display an activity cycle \citep{2012A&A...543A..84R}, and we adopt the full range of observed X-ray luminosities as the error bar for 61 Cyg A's normalized X-ray luminosity in our analysis. 61 Cyg B has been found to have a flat activity profile \citep{2012A&A...543A..84R}, and we adopt its mean X-ray luminosity and the standard deviation over all X-ray observations as the error bar for our analysis. 

\textit{CD -3710500:} This star is in a wide binary system with the white dwarf L481-60. Using the techniques outlined in section~\ref{wd_ages}, we use its reported $T_{eff}$ and $\log g$ values \citep{2012ApJ...746..144Z} to calculate an age of $1.77^{+0.65}_{-0.27}$~Gyr for L481-60, which we adopt as the age of the system. The star CD -3710500 was observed with a front-illuminated \textit{Chandra} CCD, which only provides energy sensitivity above 0.6 keV. We performed a spectral fit and extrapolated the stellar X-ray flux for the full 0.2-2 keV energy range.

\textit{GJ 176:} This star was detected with a back-illuminated \textit{Chandra} CCD, which provides energy sensitivity above 0.245 keV. We performed a spectral fit and extrapolated the stellar X-ray flux for the full 0.2-2 keV energy range.

\textit{GJ 191:}  This star was detected with a back-illuminated \textit{Chandra} CCD, which provides energy sensitivity above 0.245 keV. We performed a spectral fit and extrapolated the stellar X-ray flux for the full 0.2-2 keV energy range. Our result is consistent within errors with the values reported by \citet{Guinan2016}.

\textit{HR 7703:} This source was detected with \textit{XMM-Newton}, and we performed a spectral fit for the full energy range of 0.2-2 keV.

\textit{KIC 12011630, KIC 3123191, KIC 5309966:} These stars were in the field of view of \textit{Chandra} during observations of other targets, and are all undetected in X-rays. The sources are located on front-illuminated CCDs and are far from the centre of the field of view. Since \textit{Chandra}'s PSF becomes large at the edges of the field of view, large extraction regions had to be used, which led to quite high upper limits for the X-ray luminosities for these stars.

\textit{KIC 10016239, KIC 7529180:} These stars were detected with \textit{XMM-Newton}. KIC 7529180 was detected with a sufficient number of source counts so that a spectral fit could be performed; for KIC 10016239, the excess source counts were used to calculate the X-ray flux using a coronal temperature of $\log T = 6.5$.

\textit{KIC 6116048, KIC 6603624, KIC 8292840, KIC 9025370, KIC 9410862:} These stars were observed with \textit{XMM-Newton}, but undetected in X-rays.

\textit{KIC 9955598:} This star was observed and detected with \textit{XMM-Newton}. Since there is another X-ray source at close projected distance (ca.\ $20^{\prime\prime}$), we chose an extraction region with a radius of $10^{\prime\prime}$ instead of $20^{\prime\prime}$, and applied the correct encircled energy fraction factor to account for the smaller extraction region when calculating the flux.

\textit{NLTT 7887:} This star is in a wide binary system with the white dwarf NLTT 7890. Since the reported surface gravity of the white dwarf has large errors \citep{garces_wd}, the age we derive has large errors as well with $4.97^{+8.8}_{-3.0}$~Gyr. The star NLTT 7887 was covered by an \textit{XMM-Newton} observation, but is undetected.

\textit{$\alpha$ Cen A, $\alpha$ Cen B, Proxima Cen:} The age of this triple system has been derived from asteroseismic observations of $\alpha$ Cen A and $\alpha$ Cen B using different underlying models \citep{prox_cen_age_2}; we adopt the mean of the asteroseismic age estimates as the age of the system. $\alpha$ Cen A and $\alpha$ Cen B have been monitored in X-rays with \textit{XMM-Newton} \citep{2016arXiv161206570R}. $\alpha$ Cen B was found to display an activity cycle, and we use the full range of observed X-ray luminosities as the error bar on its X-ray luminosity for our analysis. $\alpha$ Cen A is reported by the same authors to potentially be in an activty cycle as well, however only the low-activity part has been observed so far, and there is no information on what its X-ray luminosity might be during the high-activity part of the cycle. We have therefore chosen not to include $\alpha$ Cen A in our analysis. Proxima Cen has been observed with \textit{XMM-Newton} several times as well, including multiple stellar flares; a detailed analysis is given by \citet{2011A&A...534A.133F}. We adopt their quiescent X-ray luminosity of $\log L_{\mathrm{X}} = 26.69$ for Proxima Cen in our analysis.

\clearpage
\newpage

\begin{landscape}
	\section{X-ray Luminosity and Age Results}
	\label{lx_results}
	
	\begin{minipage}{\linewidth}
		\bgroup
		\def\arraystretch{1.4}
		{
			
			\begin{tabular}{llllllll}

				Name of Star / White Dwarf & Age / Gyr & log($L_{x}$ / ergs s$^{-1}$) & log($L_{x}/R_{\odot}^{2}$ / ergs s$^{-1}$ $R_{\odot}^{-2}$) & Spectral Type & Age Determination &Age Reference\\
				\hline \hline
				16 Cyg A & $6.67^{0.81}_{0.77}$  & $26.89^{0.10}_{0.10}$ & $26.71^{0.10}_{0.10}$ & G1.5Vb & Asteroseismology & 1  \\
				16 Cyg B & $7.39^{0.89}_{0.91}$ & $<25.85$   & $<25.77$ & G3V & Asteroseismology  & 1  \\
				40 Eri A / 40 Eri B & $3.70^{3.57}_{1.34}$ & $26.81^{0.10}_{0.10}$ & $26.97^{0.10}_{0.10}$ & K0.5V & White Dwarf & 4 \\
				61 Cyg A \footnote{X-ray luminosity adopted from \citet{2012A&A...543A..84R}} & $6.00^{1.00}_{1.00}$ & $27.08^{0.23}_{0.23}$ & $27.43^{0.23}_{0.23}$ & K5Ve & Isochrone Fitting & 5 \\
				61 Cyg B \footnote{X-ray luminosity adopted from \citet{2012A&A...543A..84R}} & $6.00^{1.00}_{1.00}$ & $26.88^{0.10}_{0.10}$ & $27.33^{0.10}_{0.10}$ & K7V & Isochrone Fitting & 5 \\
				CD -3710500 / L481-60 & $1.77^{0.65}_{0.27}$ & $28.18^{0.10}_{0.10}$ & $28.22^{0.10}_{0.10}$ & G7IV & White Dwarf & 4 \\
				GJ 176 & $2.00^{0.80}_{0.80}$ & $27.03^{0.10}_{0.10}$ & $27.80^{0.10}_{0.10}$ & M2.5V & Member of HR 1614 & 6 \\
				GJ 191 & $11.00^{1.00}_{1.00}$ & $26.41^{0.10}_{0.10}$ & $27.02^{0.10}_{0.10}$ & sdM1.0 & Galatic Halo population & 7 \\
				HR 7703 & $10.00^{2.00}_{2.00}$ & $26.80^{0.10}_{0.10}$ & $27.04^{0.10}_{0.10}$ & K2.5V & Old disk star & 8 \\
				KIC 10016239 & $2.47^{0.67}_{0.61}$ & $27.91^{0.10}_{0.10}$ & $27.71^{0.10}_{0.10}$ & F6IV & Asteroseismology & 2 \\
				KIC 12011630 & $8.48^{1.53}_{1.42}$ & $<30.06$ & $<30.05$ & G1.5V & Asteroseismology & 2 \\
				KIC 3123191 & $4.26^{0.80}_{0.75}$ & $<29.12$ & $<28.84$ & F4.5V & Asteroseismology & 2 \\
				KIC 5309966 & $3.51^{1.23}_{0.42}$ & $<29.44$ & $<29.02$ & F6V & Asteroseismology & 2 \\
				KIC 6116048 & $9.58^{2.16}_{1.90}$ & $<26.89$ & $<26.74$ & F9IV-V & Asteroseismology & 1 \\
				KIC 6603624 & $7.82^{0.94}_{0.86}$ & $<27.08$ & $<26.96$ & G8IV-V & Asteroseismology & 1 \\
				KIC 7529180 & $1.93^{0.35}_{0.30}$ & $28.98^{0.10}_{0.10}$ & $28.66^{0.10}_{0.10}$ & F5IV-V & Asteroseismology & 2 \\
				KIC 8292840 & $3.85^{0.81}_{0.75}$ & $<28.14$ & $<27.88$ & F7V & Asteroseismology & 3 \\
				KIC 9025370 & $6.55^{1.26}_{1.13}$ & $<27.24$ & $<27.23$ & F8 & Asteroseismology & 1 \\
				KIC 9410862 & $6.93^{1.49}_{1.33}$ & $<27.98$ & $<27.86$ & F7V & Asteroseismology & 1 \\
				KIC 9955598 & $6.98^{0.40}_{0.50}$ & $26.90^{0.10}_{0.10}$ & $27.01^{0.10}_{0.10}$ & K0V & Asteroseismology & 3 \\
				NLTT 7887 / NLTT 7890 & $4.97^{8.80}_{3.00}$ & $<26.92$ & $<27.13$ & K2 & White Dwarf & 4 \\
				Proxima Centauri  \footnote{X-ray luminosity taken from \citet{2011A&A...534A.133F}} & $6.13^{0.55}_{0.55}$ & $26.69^{0.10}_{0.10}$ & $28.23^{0.10}_{0.10}$ & M5.5Ve & Asteroseismology & 9 \\
				Alpha Centauri B \footnote{X-ray luminosity adopted from \citet{2012A&A...543A..84R}} & $6.13^{0.55}_{0.55}$ & $27.06^{0.36}_{0.36}$ & $27.19^{0.36}_{0.36}$ & K1V & Asteroseismology & 9 \\
				Sun \footnote{X-ray luminosity adapted from model values given in \citet{2000ApJ...528..537P}} & $4.57^{0.02}_{0.02}$ & $26.77^{0.72}_{0.72}$ & $26.77^{0.72}_{0.72}$ & G2V & Isotopic Dating & 10 \\
				\hline  
     
			\end{tabular}
		}
		
		\egroup
		
		\medskip
		
		\noindent \textbf{References}: (1) \citet{SilvaAguirre2017}; (2) this work, age recalculated with BASTA algorithm \citep{astero1} using values from \citet{astero2} and  \citet{2015ApJ...808..187B}; (3) \citet{astero1}; (4) this work, see Section \ref{wd_ages}; (5) \citet{61_cyg_age}; \newline (6) \citet{gj_176_age}; (7) \citet{gj_191_age}; (8) \citet{2010_dewarf}; (9) \citet{prox_cen_age_2}; (10) \citet{1995RvMP...67..781B}
		
	\end{minipage}
	
	\clearpage
	
\end{landscape}
\newpage

\section{Stellar Properties of Sample}
\label{star_properties}
\begin{minipage}{\textwidth}
\bgroup
\def\arraystretch{1.5}
	{
	\centering
	\begin{tabular}[h]{l l l l l l}
		\hline \hline
		Name of Star / White Dwarf & Radius / $R_{\odot}$ & $T_{eff}$ / K & $V_{mag}$ & Distance / pc & References\\
		\hline 
		16 Cyg A & 1.22 & 5825 & 5.95 & 21.29 & 1,6,8,10 \\
		16 Cyg B & 1.10 & 5720 & 6.20 & 21.22 & 1,6,8,10  \\
		40 Eri A / 40 Eri B & 0.83  & 5225 & 4.43 & 4.98 & 5,6,8,9 \\
		61 Cyg A & 0.67 & 4450 & 5.21 & 3.49 & 4,6,8,9 \\
		61 Cyg B & 0.60 & 4050 & 6.03 & 3.50 & 4,6,8,9 \\
		CD -3710500 / L481-60 & 0.96 & 5530 & 6.01 & 15.25 & 5,6,8,10 \\
		GJ 176 & 0.41 & 3475 & 9.95 & 9.27 & 5,6,8,9 \\
		GJ 191  & 0.49 & 3680 & 8.85 & 3.91 & 5,6,8,9 \\
		HR 7703 & 0.77 & 4940 & 5.32 & 6.02 & 5,6,8,9 \\
		KIC 10016239 & 1.26 & 6482 & 9.81 & 175.93 & 3,7,8,10 \\
		KIC 12011630 & 1.01 & 5817 & 10.27 & 114.40 & 3,7,8,11 \\
		KIC 3123191 & 1.37 & 6568 & 9.90 & 167.67 & 3,7,8,10 \\
		KIC 5309966 & 1.62 & 6356 & 10.62 & 288.60 & 3,7,8,10 \\
		KIC 6116048 & 1.19 & 6072 & 8.47 & 75.15 & 1,7,8,10 \\
		KIC 6603624 & 1.15 & 5612 & 9.19 & 83.56 & 1,7,8,10 \\
		KIC 7529180 & 1.45 & 6682 & 8.49 & 108.53 & 3,7,8,10 \\
		KIC 8292840 & 1.35 & 6239 & 10.51 & 250.69 & 2,2,8,9 \\
		KIC 9025370 & 1.00 & 5659 & 8.95 & 87.42 & 1,7,8,10 \\
		KIC 9410862 & 1.16 & 6230 & 10.78 & 202.27 & 1,7,8,10 \\
		KIC 9955598 & 0.88 & 5434 & 9.64 & 69.39 & 2,7,8,10 \\
		NLTT 7887 / NLTT 7890 & 0.78 & 5040 & 9.84 & 41.16 & 5,6,8,10 \\
		Proxima Centauri & 0.17 & 2925 & 11.13 & 1.30 & 5,6,8,9 \\
		Alpha Centauri B & 0.86 & 5316 & 1.33 & 1.35 & 12,12,8,12 \\
		Sun  & 1.00 & 5777 & -26.74 & 1 AU & 13 \\    
		\hline 
	\end{tabular}
	}
	
	\medskip
	
	\textbf{References}: (1) \citet{SilvaAguirre2017}; (2) \citet{astero1}; (3) this work, radius recalculated with BASTA algorithm \citep{astero1} using values from \citet{astero2} and  \citet{2015ApJ...808..187B}; (4) \citet{61_cyg_age}; (5) Radius calculated from $T_{eff}$ and absolute brightness; (6) $T_{eff}$ estimated from Spectral Type using Table 5 from \citet{2013ApJS..208....9P}; (7) \citet{astero2};
	(8) $V_{mag}$ from SIMBAD; (9) Parallax from SIMBAD; (10) Parallax from \textit{Gaia} DR1 \citep{2016A&A...595A...2G}; (11) Distance from Barnes-Evans method (see Section \ref{dist_section} for details); (12) \citet{2010_dewarf}; (13) Table 1 in \citet{2010_dewarf}

\egroup
\end{minipage}

\clearpage

\section{Spectral Modelling Results}
\label{spec_model_results}

{
	\centering
	\begin{tabular}[h]{l l l l}
		\hline \hline
		Name of Star & Model kT & Model Emission Measure  & Reduced chi-squared\\
		             & (keV)          &($\frac{4\pi d^2}{10^{-14}}$ cm$^{-3}$) & \\
		\hline

		40 Eri A & 0.19 & $9.9\times 10^{-5}$ & 1.59\\ [0.3cm]
		CD -3710500 & 0.44 & $2.2\times 10^{-4}$ & 0.32\\[0.3cm]
		GJ 176 & 0.37 & $3.4\times 10^{-5}$ & 1.49\\[0.3cm]
		GJ 191 & 0.30 & $5.4\times 10^{-5}$ & 2.52\\[0.3cm]
		HR 7703 & 0.17 & $5.31\times 10^{-5}$ & 1.34\\
		& 0.76 & $1.45\times 10^{-5}$  &\\[0.3cm]
		KIC 7529180 & 0.22 & $9.8\times 10^{-6}$ & 1.70\\
		& 0.91 & $1.7\times 10^{-5}$  &\\[0.3cm]
		61 Cyg A & 0.21 & $4.09\times 10^{-4}$ & 1.99\\
		& 0.79 & $1.75\times 10^{-4}$  &\\[0.3cm]
		61 Cyg B & 0.19 & $1.71\times 10^{-4}$ & 1.17\\
		& 0.67 & $5.06\times 10^{-4}$  &\\
		\hline
	\end{tabular}
	}

\clearpage

\begin{figure*}

\begin{flushleft}
\section{Spectral Modelling Plots}
\label{spec_model_plots}
\end{flushleft}

\begin{subfigure}{1.0\columnwidth}
	\centering
	\includegraphics[height = 0.2\paperheight,width=0.9\columnwidth]{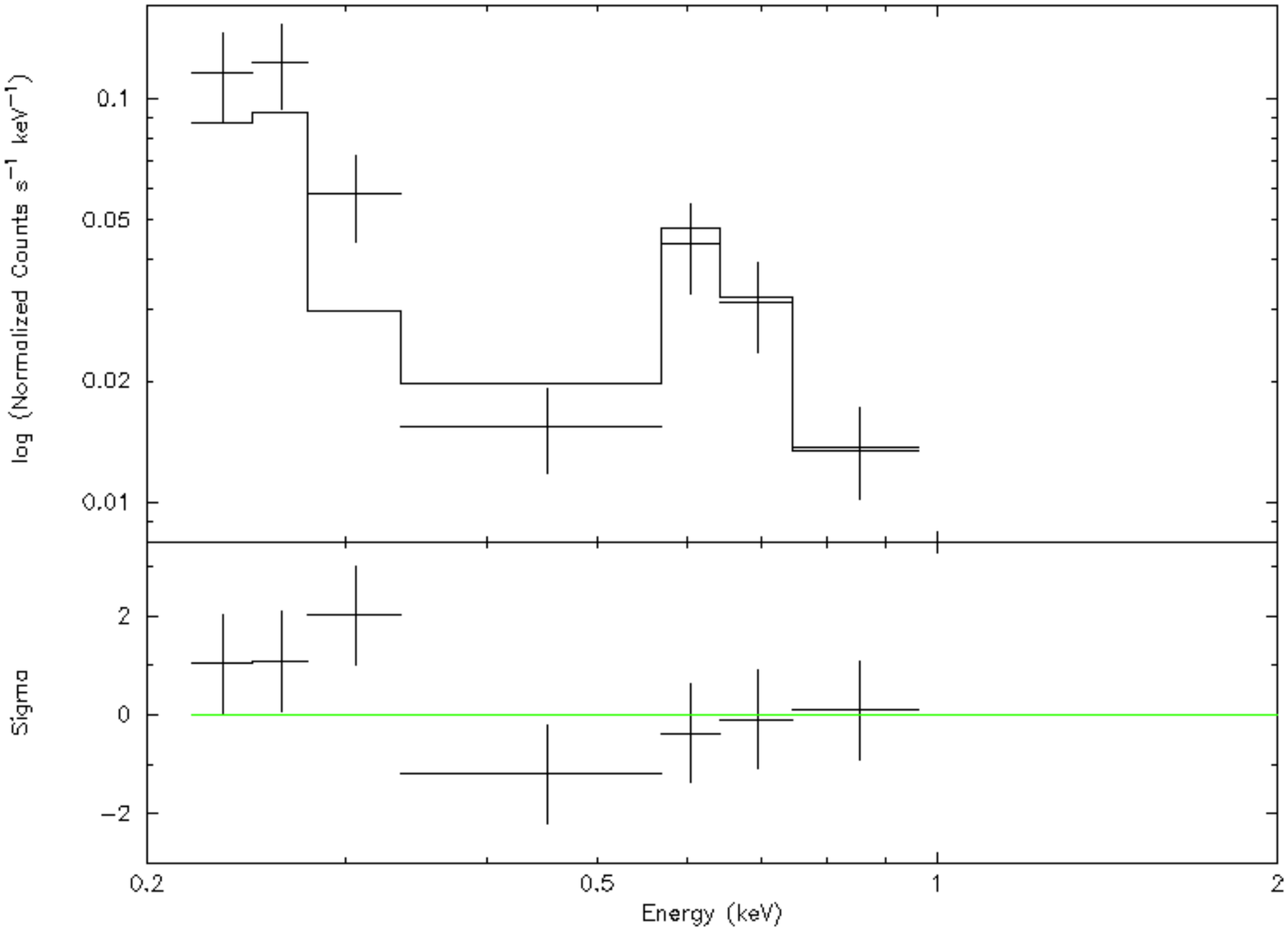}
	\caption{40 Eri A}
\end{subfigure}
\begin{subfigure}{1.0\columnwidth}
	\centering
	\includegraphics[height = 0.2\paperheight,width=0.9\columnwidth]{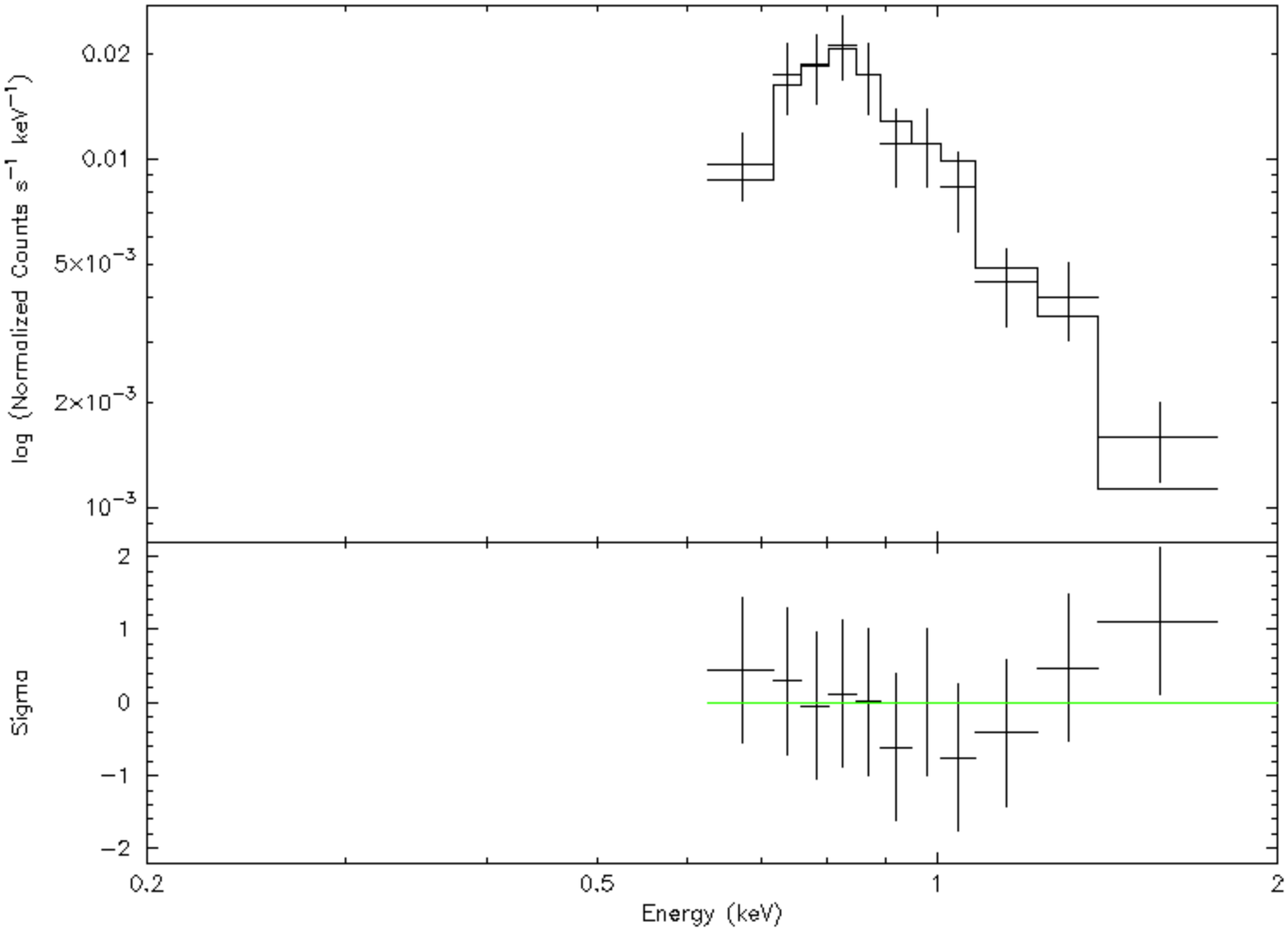}
	\caption{CD -3710500}
\end{subfigure}
\hfill{}
\begin{subfigure}{1.0\columnwidth}
	\centering
	\includegraphics[height = 0.2\paperheight,width=0.9\columnwidth]{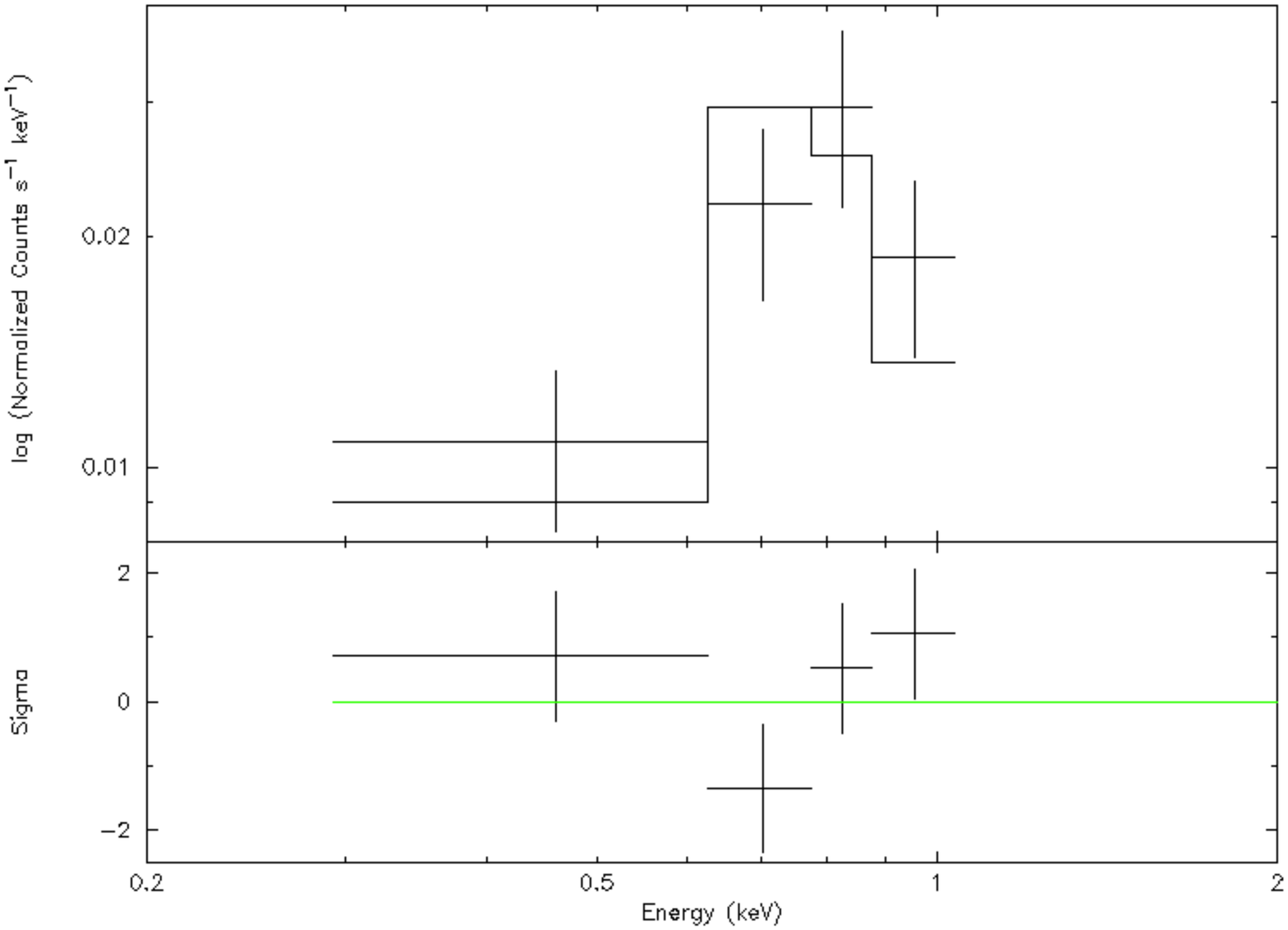}
	\caption{GJ 176}
\end{subfigure}
\begin{subfigure}{1.0\columnwidth}
	\centering
	\includegraphics[height = 0.2\paperheight,width=0.9\columnwidth]{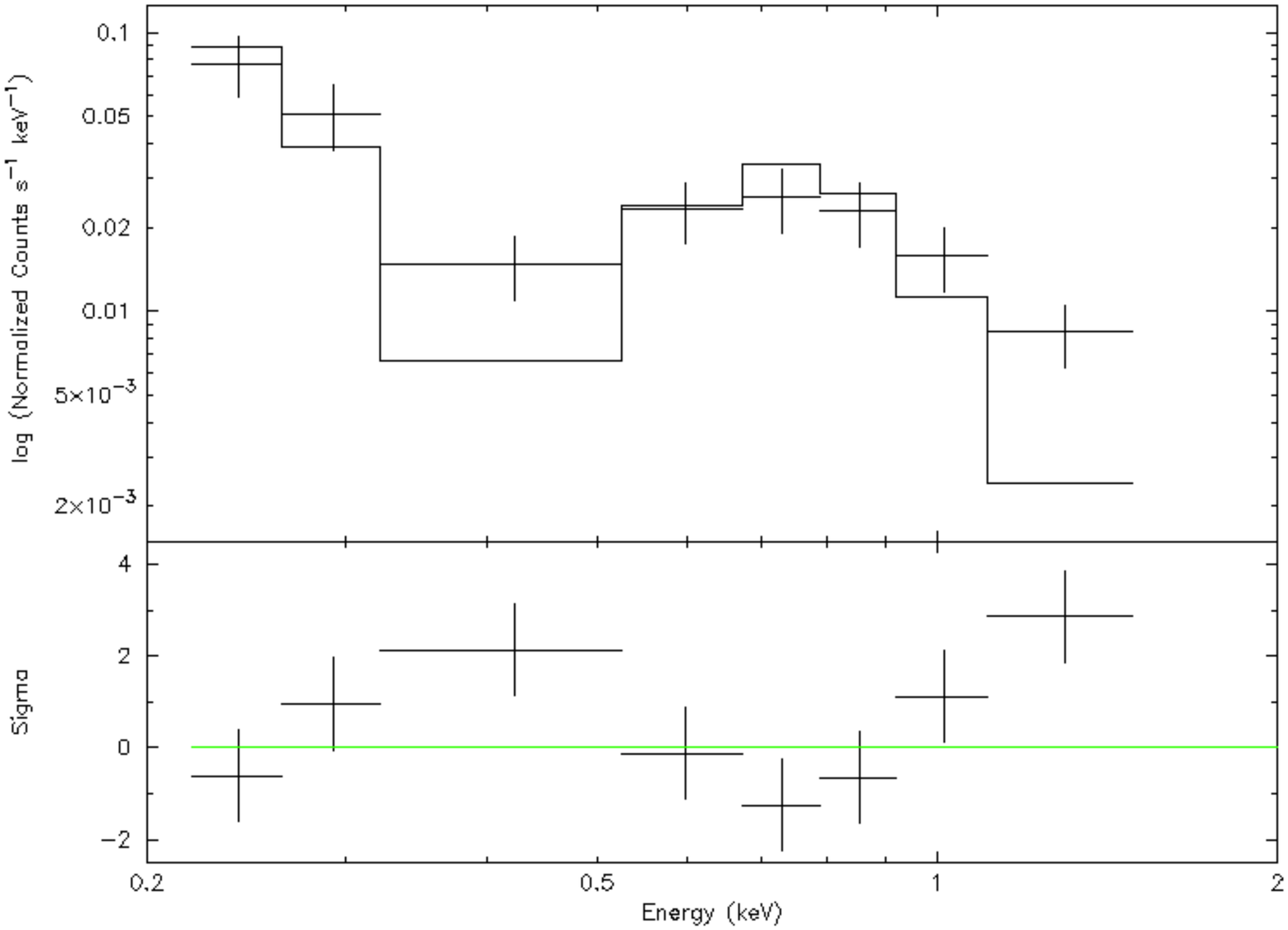}
	\caption{GJ 191}
\end{subfigure}
\hfill{}
\begin{subfigure}{1.0\columnwidth}
	\centering
	\includegraphics[height = 0.2\paperheight,width=0.9\columnwidth]{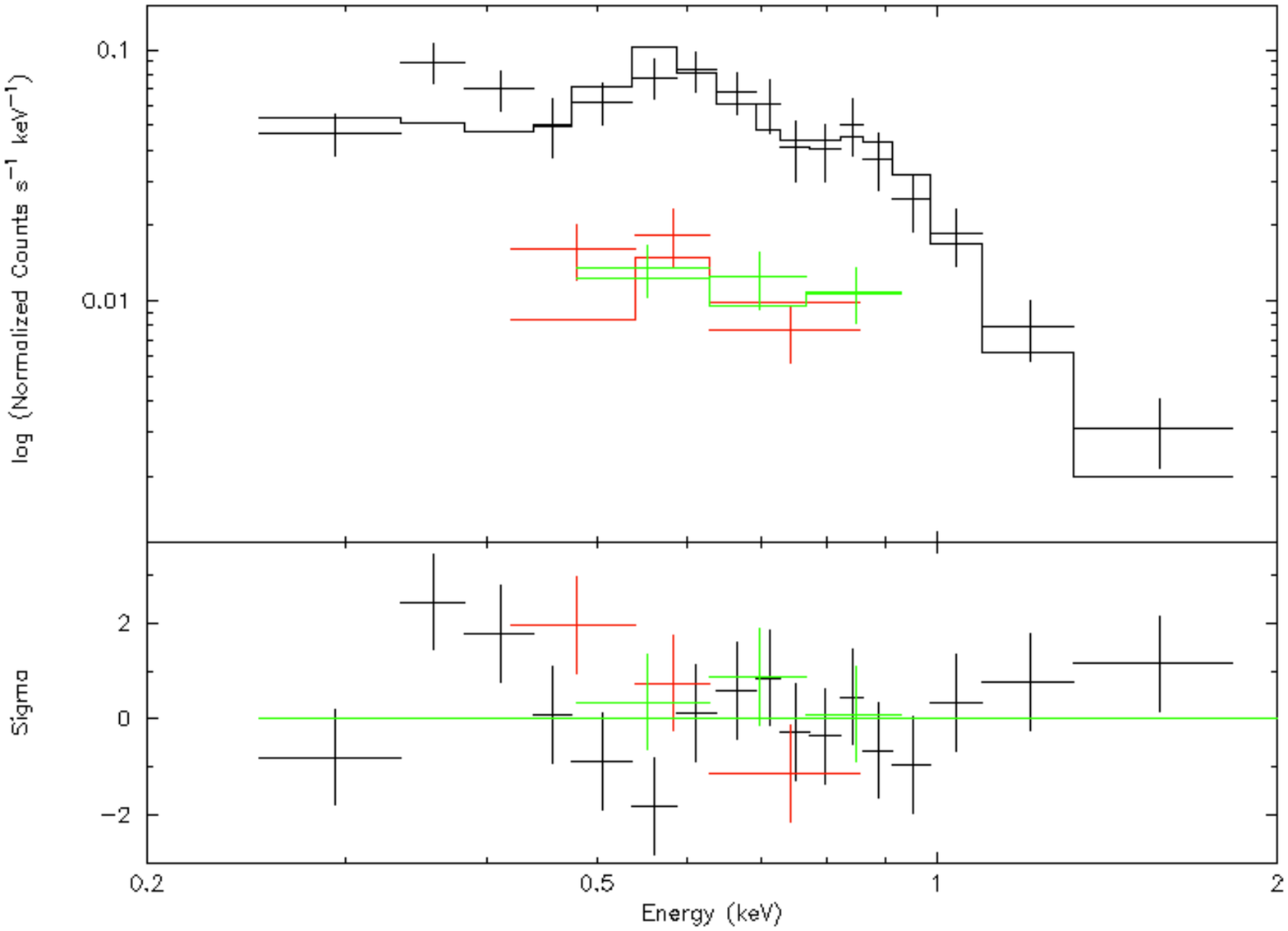}
	\caption{HR 7703}
\end{subfigure}
\begin{subfigure}{1.0\columnwidth}
	\centering
	\includegraphics[height = 0.2\paperheight,width=0.9\columnwidth]{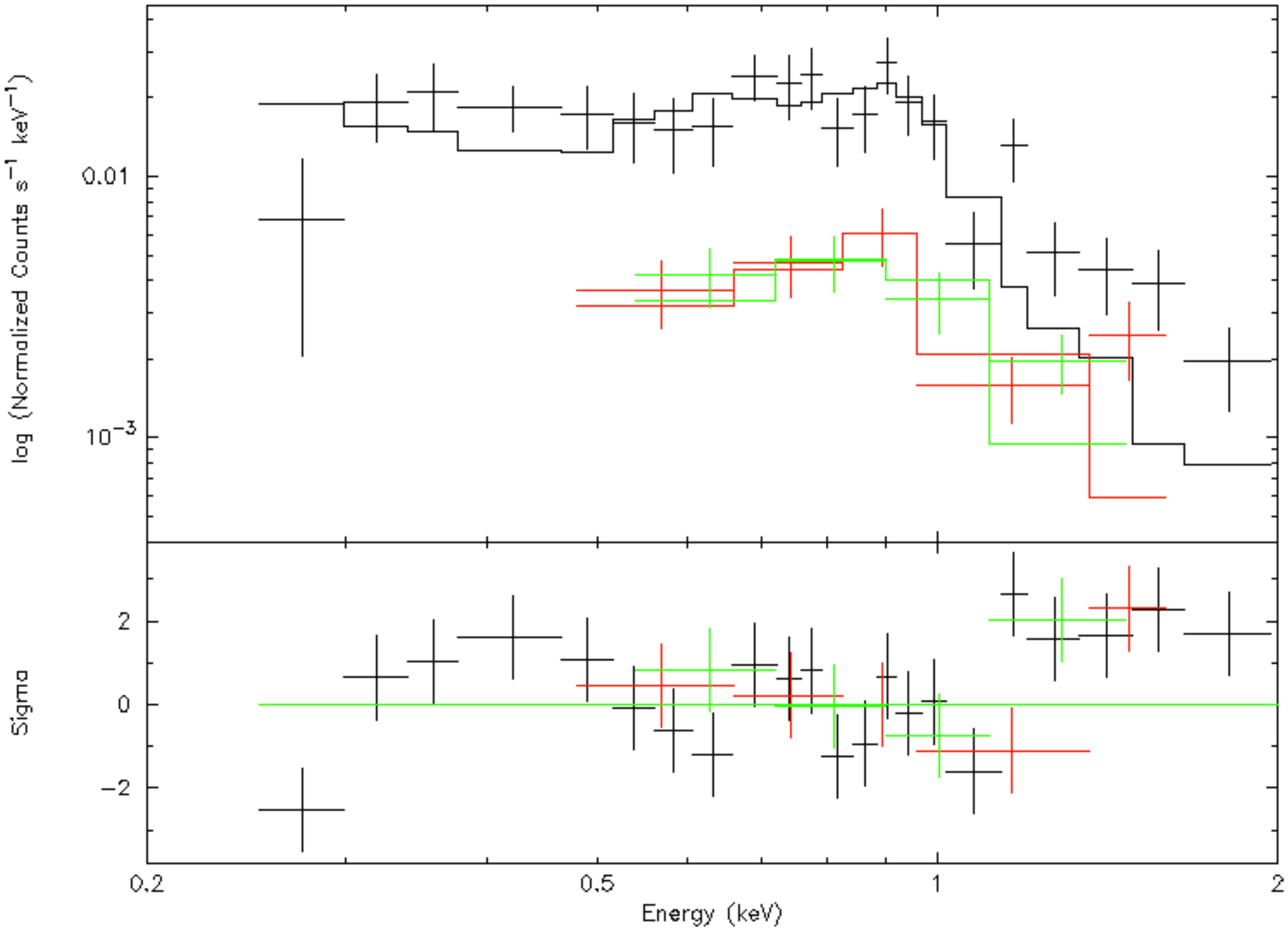}
	\caption{KIC 7529180}
\end{subfigure}
\par\bigskip

\caption{X-ray spectra (grouped to 15 counts per bin) and best fit models of six X-ray sources. These sources were detected to at least three sigma and contained over a ninety counts in the source region. The top region of each subplot shows the number of counts per second per keV as a function of energy. The bottom region of each subplot shows the sigma value for the best fit model as a function of energy. Different colours indicate spectra from different detectors which are fitted simultaneously to ensure a more accurate fit. CD -3710500 was observed with a front-illuminated \textit{Chandra} CCD and therefore only has spectral data above 0.6 keV.}

\end{figure*}

\end{appendices}

\end{document}